\documentclass[aps,preprint,amsmath,amssymb,amsfonts]{revtex4}
\usepackage{epsfig}
\usepackage{graphicx}
\usepackage{subfigure}
\usepackage{dcolumn}
\usepackage{bm}
\usepackage{amsthm}
\usepackage{enumitem}
\usepackage{slashed}
\usepackage{braket}
\usepackage{amsmath}
\usepackage{mathtools}
\usepackage{bbold}

\usepackage{color}   
\usepackage{hyperref}
\hypersetup{
	colorlinks=true,  
	linkcolor=black,  
	citecolor=black,
	filecolor=black,
	urlcolor=black,
}

\makeatletter
\def\l@subsubsection#1#2{}
\makeatother

\newcommand{\del}{\partial}

\newcommand{\sign}{\operatorname{sign}}

\newcommand{\bbR}{\mathbb{R}}

\begin{document}

\title{Self-dual gravity in de Sitter space: lightcone ansatz and static-patch scattering}

\author{Yasha Neiman}
\email{yashula@icloud.com}
\affiliation{Okinawa Institute of Science and Technology, 1919-1 Tancha, Onna-son, Okinawa 904-0495, Japan}

\date{\today}

\begin{abstract}
Using Krasnov's formulation of General Relativity (GR), we develop a lightcone ansatz for self-dual gravity (along with linearized anti-self-dual perturbations) in the Poincare patch of de Sitter space. This amounts to a generalization of Plebanski's ``second heavenly equation'' to non-zero cosmological constant. The only interaction vertices are cubic ones, found previously by Metsaev in a bottom-up lightcone approach. We point out a special feature of these vertices, which leads to ``almost conservation'' of energy at each successive order in perturbation theory, despite the time-dependent de Sitter background. Since we embed the lightcone variables into a full spacetime metric, the solutions have a clear geometric interpretation. In particular, this allows us to read off boundary data on both the past and future horizons of a causal (static) patch. In this way, we add self-dual GR to the program of defining \& computing scattering amplitudes in a causal patch of de Sitter space.
\end{abstract}

\maketitle
\tableofcontents
\newpage

\section{Introduction and structure of the paper} \label{sec:intro}

In this paper, we find an exact ansatz for self-dual GR solutions (as well as linearized anti-self-dual perturbations on top of these) over a de Sitter background, using a lightcone gauge in Poincare coordinates. While the ansatz can be described in standard metric language, in its derivation we use Krasnov's chiral, metric-free reformulation of GR \cite{Krasnov:2011pp,Krasnov:2011up}. Our lightcone ansatz reduces each helicity of the gravitational field to a scalar degree of freedom (as usual in lightcone formulations). The dynamics of these scalars is then given by an action with only cubic vertices, which leads to field equations that can be solved perturbatively. Such a cubic action has been worked out previously by Metsaev \cite{Metsaev:2018xip}, in the context of describing the most general cubic interactions for massless fields in (A)dS\textsubscript{4} in a lightcone formalism. The novelty in our approach is that we show how the lightcone degrees of freedom are actually related to components of the spacetime metric and curvature.

Another important historical comparison is to Plebanski's ``heavenly equations'' \cite{Plebanski:1975wn}, which describe solutions of self-dual GR with zero cosmological constant $\Lambda=0$ in terms of a scalar degree of freedom. Plebanski found two such equations, which provide alternative descriptions of self-dual solutions. Both are quadratic in the scalar variable, corresponding to cubic-only vertices. The first of the ``heavenly equations'' was extended to $\Lambda\neq 0$ by Przanowski in the 1980's \cite{Przanowski:1983xpa}. In this extension, the equation goes from quadratic to non-polynomial, i.e. we get interaction vertices of all orders. The present paper's lightcone ansatz constitutes the (missing up to now) $\Lambda\neq 0$ extension of the \emph{second} ``heavenly equation''. Remarkably, this extension \emph{does} preserve the equation's quadratic nature. \{Note added: shortly after the first version of the present paper, the same extension was achieved independently in \cite{Lipstein:2023pih}.\}

Our motivation for describing this ansatz for self-dual GR is to add self-dual GR to the list of theories for which we can define \& compute \emph{scattering in the de Sitter static patch} -- the closest thing to an ``asymptotic observable'' that's available to an observer inside de Sitter space. As we describe below, for full GR, already defining this scattering problem is difficult, even at the classical level. However, in the self-dual sector, we find that the scattering problem can be addressed by methods similar to those we used in \cite{Albrychiewicz:2021ndv} for Yang-Mills theory.

Since our lightcone-gauge results should be of interest also outside the context of static-patch scattering, the paper has a somewhat hybrid structure. In section \ref{sec:static}, we motivate and pose the static-patch scattering problem. In section \ref{sec:ansatz}, we describe the lightcone ansatz, sketch the resulting perturbative framework for tree-level computations, and note that it possesses the curious feature of ``almost conserved'' energy (despite the absence of time-translation symmetry in Poincare coordinates). In section \ref{sec:Krasnov}, we review Krasnov's formulation of GR, and use it to derive our ansatz and its cubic action. In section \ref{sec:together}, we return to the static-patch problem, and show how it's addressed by our Poincare-patch, lightcone-gauge solutions (as a general reference on the relationship between tree-level scattering and classical field solutions, see e.g. \cite{Boulware:1968zz}). In section \ref{sec:outlook}, we discuss directions for future work. In an Appendix, we address the question of our lightcone ansatz's generality, showing that any perturbation away from it is (at least locally) a gauge transformation.

Throughout the paper, we use ``right-handed/left-handed'' as synonymous with ``self-dual/anti-self-dual'', respectively.

\section{Scattering in the static patch} \label{sec:static}

\subsection{Motivation} \label{sec:static:motivation}

In theoretical physics, we are generally more comfortable with problems posed at spacetime's asymptotic boundary at infinity. This includes scattering amplitudes in Minkowski space, as well as boundary correlators in AdS/CFT. Moreover, in quantum gravity, there is a general expectation that sharp observables can \emph{only} be defined at infinity, since that is where we have a fixed reference geometry. On the other hand, our world seems to have a positive cosmological constant $\Lambda>0$. This traps any physical observers inside a cosmological horizon of finite size, and challenges us to confront quantum field theory and gravity in finite regions after all. 

The simplest setup in which to approach these issues is the \emph{static patch of de Sitter space}, i.e. the largest causally-connected region of the simplest spacetime with positive $\Lambda$ (see e.g. \cite{Bousso:2000nf,Anninos:2012qw,Halpern:2015zia}). The static patch is a region enclosed by a pair of cosmological horizons: one in the future, and one in the past. The past (future) horizon is the lightcone of a point at past (future) asymptotic infinity. In other words, the static patch is the largest \emph{causal diamond} in de Sitter space. Taking a cue from the better-understood observables at Minkowski or AdS infinity, we can now try and compute observables on the static patch's boundaries. This leads to the idea of \emph{scattering} in the static patch, i.e. computing the evolution of suitable field data from the past horizon to the future one \cite{David:2019mos}. Note that this is quite different from the more standard problem considered in de Sitter space, namely of correlators at the future conformal boundary. The latter only become observable in a universe that exits its de Sitter phase (as is conjectured to be the case with inflation). 

The static-patch scattering problem has so far been considered for free massless fields of arbitrary spin \cite{David:2019mos}, for a scalar with cubic interaction \cite{Albrychiewicz:2020ruh}, and (with much greater success) for Yang-Mills theory \cite{Albrychiewicz:2021ndv} -- all at tree level, i.e. at the level of classical field solutions. Our present goal is to add GR to this list. In general, this is a difficult problem already conceptually, since gravitational perturbations will alter the causal structure of the static patch itself and its horizon boundaries. Our compromise will be to consider GR's \emph{self-dual sector}, i.e. the sector with vanishing left-handed Weyl curvature. Within this sector, as we'll see shortly, the causal structure remains sufficiently intact: the horizon maintains the same constant area all the way to conformal infinity, neither collapsing into caustics nor expanding to reach causally-disconnected points at the conformal boundary.

The price we pay is that self-dual metrics (in Lorentzian signature) are necessarily complex, and thus do not describe physically meaningful geometries. Of course, in Euclidean signature, self-dual solutions can be real, and their study is a staple of the GR literature. Such Euclidean solutions are often called gravitational instantons, and considered as tunneling-type contributions to the gravitational path integral \cite{Hawking:1976jb}. Computations in the self-dual sector are also useful ``directly'' in Lorentzian signature. For instance, in Minkowski space, the self-dual sectors of Yang-Mills and GR form the structure behind MHV amplitudes, which are the starting point for all other amplitude calculations \cite{Bardeen:1995gk,Rosly:1996vr,Mason:2009afn} (for a striking example, see the explanation \cite{Bern:2017puu} of GR's 2-loop divergence in terms of a 1-loop anomaly in the self-dual sector). The general point here is that complex classical solutions become physically meaningful at the quantum level. In our static-patch case, the technical simplicity of the self-dual sector is accompanied by a greater conceptual clarity (due to the horizons maintaining their causal structure), again hopefully making it a valuable starting point for future explorations. 

Note that, for the more standard inflationary-correlators problem, the self-dual sector is not sufficient to compute \emph{any} correlators. This is because the bulk self-duality condition is inconsistent with the standard boundary conditions at future conformal infinity. For the static-patch problem, this isn't the case: as we saw in \cite{Albrychiewicz:2021ndv} for Yang-Mills, the self-dual sector precisely computes the static-patch analog of MHV (more precisely, N\textsuperscript{$-1$}MHV) amplitudes. The situation for self-dual GR is the same.

Despite our ultimate interest in the static patch, we perform most of the calculation in Poincare coordinates. This ``trick'' considerably simplifies our task, thanks to the higher symmetry of Poincare coordinates, especially spatial translation invariance. Thus, we set up our calculation in a Poincare patch whose past boundary coincides with our static patch's past horizon. Then, at the very end, we read off the field data on the static patch's future horizon, which, from the point of view of the Poincare patch, is just some lightlike bulk hypersurface. The price or this ``cheating''  is that we need to solve the Poincare-patch evolution not just in one gauge, but in \emph{every lightcone gauge}: to read off the final data on each separate lightray of the future horizon, we need a lightcone gauge adapted to that lightray (see section \ref{sec:together}, as well as the Yang-Mills version \cite{Albrychiewicz:2021ndv}).

\subsection{Framework} \label{sec:static:framework}

Let us now formulate the static-patch problem more concretely. We will need to be more careful than in previous treatments \cite{David:2019mos,Albrychiewicz:2020ruh,Albrychiewicz:2021ndv}, which did not involve a dynamical geometry. In particular, in \cite{David:2019mos,Albrychiewicz:2020ruh,Albrychiewicz:2021ndv} we only considered the 3d initial/final data on the past/future horizon, while ignoring any extra data or constraints on the horizons' 2d intersection. In the present context, some constraints on the 2d intersection will prove important for a proper geometric and causal interpretation of our scattering problem.

\subsubsection{Geometric assumptions on the horizons' intersection} \label{sec:static:framework:intersection}

In pure $dS_4$, the past and future horizons of the static patch intersect at a 2-sphere whose intrinsic curvature radius equals that of $dS_4$ itself. From here on, we set this curvature radius to 1. The \emph{extrinsic} curvature of this 2-sphere vanishes, i.e. the covariant derivative along the sphere of its two lightlike normals is zero. In our static-patch scattering problem, we will assume that this intrinsic and extrinsic geometry of the intersection 2-sphere remains undeformed.

To express these properties in equations, let us denote points on the 2-sphere as unit 3d vectors $\mathbf{\hat r}$, and the lightlike coordinates orthogonal to the 2-sphere by $u,v$, such that the 2-sphere is at $u=v=0$. We take the $u$ coordinate to be future-pointing, and $v$ past-pointing. We define basis vectors with respect to our coordinates as:
\begin{align}
  \ell \equiv \frac{\del}{\del u} \ ; \quad n \equiv \frac{\del}{\del v} \ ; \quad m \equiv \mathbf{m}\cdot\frac{\del}{\del\mathbf{\hat r}} \ ; \quad \bar m \equiv \mathbf{\bar m}\cdot\frac{\del}{\del\mathbf{\hat r}} \ , \label{eq:basis}
\end{align}
where:
\begin{align}
 \mathbf{\hat r\cdot\hat r} = 1 \ ; \quad \mathbf{m\cdot\hat r} = \mathbf{\bar m\cdot\hat r} = \mathbf{m\cdot m} = \mathbf{\bar m\cdot\bar m} = 0 \ ; \quad \mathbf{m\cdot\bar m} = \frac{1}{2} \ .
\end{align}
We fix the inner product of the lightlike normals as $g_{\ell n}=\frac{1}{2}$, and we take the bivector $\ell\wedge m$ to be left-handed. The 2-sphere's intrinsic and extrinsic geometry can now be expressed by the following relations at $u=v=0$:
\begin{gather}
   g_{\ell\ell} = g_{nn} = g_{\ell m} = g_{\ell\bar m} = g_{nm} = g_{n\bar m} = g_{mm} = g_{\bar m\bar m} = 0 \ ; \quad g_{\ell n} = g_{m\bar m} = \frac{1}{2} \ ; \label{eq:intersection_metric} \\
    \del_u g_{mm} = \del_u g_{\bar m\bar m} = \del_u g_{m\bar m} = \del_u g_{nm} = \del_u g_{n\bar m} = 0 \ ; \label{eq:intersection_u_derivs} \\
    \del_v g_{mm} = \del_v g_{\bar m\bar m} = \del_v g_{m\bar m} = \del_v g_{\ell m} = \del_v g_{\ell\bar m} = 0 \ . \label{eq:intersection_v_derivs}
\end{gather}
We note that these properties fix the 2-sphere's coordinates and the basis vectors \eqref{eq:basis}, up to a $SO(1,1)\times SO(3)$ global symmetry and a $U(1)$ local symmetry. Here, the $SO(3)$ are spatial rotations of the $\mathbf{\hat r}$ coordinates, the $SO(1,1)$ are rescalings of $u$ and $v$ (or, equivalently, $\ell$ and $n$) by equal and opposite factors, and the $U(1)$ are phase rotations (or, equivalently, $SO(2)$ geometric rotations around $\mathbf{\hat r}$) of the basis vectors $\mathbf{m},\mathbf{\bar m}$ at each point $\mathbf{\hat r}$. Note that the global $SO(1,1)\times SO(3)$ is precisely the isometry group of the static patch in pure $dS_4$ (the $SO(1,1)$ being the static-patch time translations), while the local $U(1)$ is the standard ``little group'' of helicity transformations.

\subsubsection{Geometry and dynamical data on the past and future horizons} \label{sec:static:framework:horizons}

With the above assumptions on the intersection 2-sphere, we can now draw e.g. the future horizon by parallel-transporting the lightlike normal $\ell\equiv \del_u$ along itself, while keeping the $v$ coordinate fixed. The future horizon is then the lightlike hypersurface $v=0$, coordinatized by $(u,\mathbf{\hat r})$ with $u\geq 0$, where the lightrays are the lines of constant $\mathbf{\hat r}$, and $u$ is an affine coordinate along them. These properties can be encoded by the following relations at $v=0$:
\begin{align}
    g_{\ell\ell} = g_{\ell m} = g_{\ell\bar m} = 0 \ ; \quad \del_v g_{\ell\ell} = 0 \ ; \quad g_{\ell n} = \frac{1}{2} \ , \label{eq:future_horizon_metric_1}
\end{align}
where the last two conditions ensure the affine nature of the lightlike vector $\ell$ (and its associated coordinate $u$). Similarly, we draw the past horizon at $(u=0,v\geq 0)$, with:
\begin{align}
    g_{nn} = g_{nm} = g_{n\bar m} = 0 \ ; \quad \del_u g_{nn} = 0 \ ; \quad g_{\ell n} = \frac{1}{2} \label{eq:past_horizon_metric}  
\end{align}
We now come to a key property of self-dual GR. On a solution of self-dual GR, our assumptions \eqref{eq:intersection_metric}-\eqref{eq:intersection_v_derivs} vis. the intrinsic and extrinsic geometry at $u=v=0$ have consequences that hold throughout the horizons. Specifically, on e.g. the $v=0$ future horizon, in addition to \eqref{eq:future_horizon_metric_1}, we have:
\begin{align}
    g_{mm} = 0 \ ; \quad g_{m\bar m} = \frac{1}{2} \ . \label{eq:future_horizon_metric_2}
\end{align}
On other words, the only element of the horizon metric that \emph{is} deformed from its pure-$dS_4$ value is $g_{\bar m\bar m}$. Eqs. \eqref{eq:future_horizon_metric_2} imply that the horizon's expansion rate vanishes $\theta = 0$, as does its left-handed shear $\sigma_{mm} = 0$, with only the right-handed shear $\sigma_{\bar m\bar m} = \del_u g_{\bar m\bar m}$ non-vanishing. To see that \eqref{eq:future_horizon_metric_2} indeed follows from our assumptions and from self-dual GR, note that:
\begin{itemize}
    \item The initial values $g_{mm} = 0$, $g_{m\bar m} = \frac{1}{2}$ and first derivatives $\sigma_{mm} = \del_u g_{mm} = 0$, $\theta = 2\del_u g_{m\bar m} = 0$ at $u=0$ are fixed by the assumptions \eqref{eq:intersection_metric}-\eqref{eq:intersection_u_derivs}.
    \item The second derivative $\del_u \sigma_{mm}$ is governed by the product $\theta\sigma_{mm}$ and the left-handed Weyl curvature element $C_{\ell m\ell m}$, which vanishes on solutions of self-dual GR.
    \item The second derivative $\del_u\theta$ is governed by the products $\theta^2,\sigma_{mm}\sigma_{\bar m\bar m}$ and the traceless Ricci curvature element $R_{\ell\ell}$, which vanishes on vacuum solutions of GR.
\end{itemize}
In this way, the field equations of self-dual GR fix the metric elements \eqref{eq:future_horizon_metric_2} throughout the horizon, by propagating the initial conditions $\sigma_{mm}=\theta=0$ from the intersection 2-sphere.

Furthermore, the non-trivial metric element $g_{\bar m\bar m}$ on the future horizon $v=0$ can now be simply related to a right-handed Weyl curvature element $C_{\ell\bar m\ell\bar m} = \del_u^2 g_{\bar m\bar m}$. Fixing $C_{\ell\bar m\ell\bar m}$ on the horizon is equivalent to fixing $g_{\bar m\bar m}$, since eqs. \eqref{eq:intersection_metric}-\eqref{eq:intersection_u_derivs} set $g_{\bar m\bar m}=\del_u g_{\bar m\bar m} = 0$ at $u=0$. 

Finally, in addition to the right-handed horizon data $C_{\ell\bar m\ell\bar m}(u,\mathbf{\hat r})$, we consider \emph{linearized} left-handed data: the component $c_{\ell m\ell m}(u,\mathbf{\hat r})$ of a linearized left-handed Weyl curvature perturbation. Unlike the right-handed Weyl tensor $C$, we treat this $c$ not as \emph{derived from} the deformed metric, but merely as \emph{propagating on top of it}.

The same remarks apply to the \emph{past} horizon, with the replacements:
\begin{align}
    u\leftrightarrow v \ ; \quad \left(\ell \equiv \frac{\del}{\del u}\right) \leftrightarrow \left(n \equiv \frac{\del}{\del v}\right) \ ; \quad m\leftrightarrow\bar m \ .
\end{align}

\subsubsection{The scattering problem} \label{sec:static:framework:scattering}

The horizons' vanishing expansion $\theta=0$ has important consequences for their causal structure:
\begin{itemize}
    \item The horizons continue to infinite values of affine lightlike time ($v\rightarrow\infty$ for the past horizon, $u\rightarrow\infty$ for the future horizon).
    \item The area density remains constant along the horizons' lightrays.
\end{itemize}
This means that our deformed static patch remains bounded by lightlike horizons of constant finite area, just like in pure $dS_4$. It therefore remains the ``largest'' causally closed region in the deformed spacetime, which keeps us faithful to the motivation of section \ref{sec:static:motivation}. We stress again that this is a special feature of self-dual GR, where we can have non-trivial right-handed shear ($\sigma_{mm}$ or $\sigma_{\bar m\bar m}$ on the past/future horizon respectively), but vanishing left-handed shear ($\sigma_{\bar m\bar m}$ or $\sigma_{mm}$ respectively). In any \emph{real} GR solution other than pure $dS_4$, the horizons wouldn't maintain their constant area, due to the nonzero gravitational-wave energy density $\sim \sigma_{mm}\sigma_{\bar m\bar m}$.

With this understanding, we define our (classical) static-patch scattering problem as computing the data $C_{\ell\bar m\ell\bar m}(u,\mathbf{\hat r})$ and $c_{\ell m\ell m}(u,\mathbf{\hat r})$ on the future horizon from (the self-dual GR solution defined by) their counterparts $C_{nmnm}(v,\mathbf{\hat r})$ and $c_{n\bar mn\bar m}(v,\mathbf{\hat r})$ on the past horizon.

One can now ask whether this scattering problem is well-defined. Specifically:
\begin{enumerate}
    \item Given our initial data \& assumptions on the past horizon, is there a self-dual GR solution whose future horizon satisfies our extrinsic curvature constraints $(m\cdot\nabla)\ell = (\bar m\cdot\nabla)\ell = 0$ at the intersection 2-sphere?
    \item Is this solution unique?
\end{enumerate}
At the coarse-grained level of only considering 3d data, i.e. functions of all 3 coordinates on the past/future horizon, it's easy to guess that the answer is Yes, since:
\begin{itemize}
    \item The Weyl curvature components that we defined as our initial/final data are the standard ones for the characteristic-value problem in GR, and in massless theories more generally \cite{Sachs:1962zzb,Penrose:1980yx,Reisenberger:2007ku,David:2019mos}.
    \item Self-dual GR should be a self-contained sector, i.e. initial data consistent with self-duality should lead to a self-dual solution.
\end{itemize}
However, at the more fine-grained level of lower-dimensional data (i.e. 2d data at the intersection 2-sphere), the answer is less obvious. In section \ref{sec:together}, we will constructively demonstrate existence, by building initial data \eqref{eq:initial_phi} for a solution in the lightcone ansatz \eqref{eq:g_ansatz}. As for uniqueness, it follows in 3 steps:
\begin{enumerate}
    \item In the Appendix, we will show that the lightcone ansatz \eqref{eq:g_ansatz} is (locally) the most general solution to self-dual GR, up to gauge and diffeomorphisms.
    \item From the construction in section \ref{sec:together}, it will be clear that the initial data \eqref{eq:initial_phi} for the ansatz \eqref{eq:g_ansatz} is uniquely determined by (a) the curvature data $C_{nmnm}(v,\mathbf{\hat r})$ on the past horizon and (b) the constraints \eqref{eq:intersection_metric}-\eqref{eq:intersection_v_derivs} on the intersection 2-sphere. 
    \item The field equation \eqref{eq:phi_equation} governing the ansatz \eqref{eq:g_ansatz} has a hyperbolic linear term, i.e. it is a hyperbolic equation at each order in perturbation theory. This guarantees that, at least perturbatively, a solution is uniquely specified by the initial data at $v\geq 0$ as provided by \eqref{eq:initial_phi}.
\end{enumerate}

\subsubsection{Reference example: pure de Sitter space} \label{sec:static:framework:pure}

It will be useful to have explicit formulas for the $(u,v,\mathbf{\hat r})$ coordinates in pure de Sitter space $dS_4$. To achieve this, we consider flat 5d spacetime $\bbR^{1,4}$, parameterized by ``lightcone'' coordinates $(u,v,\mathbf{r})$ with metric $ds^2 = du dv + \mathbf{dr}^2$. Here, $u$ and $v$ are lightlike coordinates, and $\mathbf{r}\in\bbR^3$ is an ordinary Euclidean vector. De Sitter space $dS_4$ is then the hyperboloid $uv+\mathbf{r}^2 = 1$ within $\bbR^{1,4}$. Defining $\mathbf{\hat r}\equiv\mathbf{r}/|\mathbf{r}|$, we obtain the desired $(u,v,\mathbf{\hat r})$ coordinates for $dS_4$. The metric in these coordinates reads:
\begin{align}
    ds^2 = \frac{v^2du^2 + 2(2-uv)dudv + u^2dv^2}{4(1-uv)} + (1-uv)\mathbf{d\hat r\cdot d\hat r} \ , \label{eq:uv}
\end{align}
where $\mathbf{d\hat r\cdot d\hat r}$ is the metric of the unit 2-sphere, and the $u,v$ coordinate range is restricted by $uv\leq 1$. The hypersurfaces $u=0$ and $v=0$ are lightlike horizons that define a static patch, and have all the properties described in sections \ref{sec:static:framework:intersection}-\ref{sec:static:framework:horizons}. Of course, in pure $dS_4$, both the initial and final data for the Weyl curvature vanish.

\section{The lightcone-gauge ansatz and its perturbation theory} \label{sec:ansatz}

We now forget temporarily about static-patch scattering, and turn to present our lightcone-gauge ansatz for self-dual GR solutions in the Poincare patch. The present section merely describes the ansatz and its associated perturbation theory. In section \ref{sec:Krasnov}, we derive the ansatz's validity using the Krasnov formulation of GR. In section \ref{sec:together}, we tie it back to the static-patch problem. In the Appendix, we discuss the ansatz's generality up to gauge and diffeomorphisms.

\subsection{Poincare coordinates and spinors}

We use Poincare coordinates $x^a = (t,\mathbf{x})$ for pure $dS_4$, with metric:
\begin{align}
    ds^2 = \frac{1}{t^2}\eta_{ab}dx^a dx^b = \frac{1}{t^2}(-dt^2 + \mathbf{dx}^2) \ , \label{eq:Poincare}
\end{align}
where $\eta_{ab}$ is the flat 4d Minkowski metric. The Poincare coordinates $x^a$ are related to the embedding-space coordinates of section \ref{sec:static:framework:pure} via:
\begin{align}
    (u,v,\mathbf{r}) = -\frac{1}{t}(1, t^2 - \mathbf{x}^2, \mathbf{x}) \ . \label{eq:uv_from_Poincare}
\end{align}
The $t$ coordinate ranges from $t=-\infty$ (the past boundary of the Poincare patch) to $t=0$ (the conformal future boundary of $dS_4$). It will be very convenient to introduce spinor indices for the ``flat'' Poincare coordinates $x^a$, via the Pauli matrices $\sigma_a^{\alpha\dot\alpha}$:
\begin{align}
    x^{\alpha\dot\alpha} \equiv \sigma_a^{\alpha\dot\alpha} x^a \ ; \quad \eta_{ab}x^a x^b = -\frac{1}{2}x_{\alpha\dot\alpha}x^{\alpha\dot\alpha} \ ; \quad
    \del^{\alpha\dot\alpha} \equiv \eta^{ab}\sigma_a^{\alpha\dot\alpha}\frac{\del}{\del x^b} = -2\frac{\del}{\del x_{\alpha\dot\alpha}} \ . \label{eq:spinors}
\end{align}
The left-handed spinor indices $(\alpha,\beta,\dots)$ are raised and lowered with the flat antisymmetric spinor metric $\epsilon_{\alpha\beta}$, and likewise for the right-handed indices $(\dot\alpha,\dot\beta,\dots)$:
\begin{align}
    \zeta_\alpha = \epsilon_{\alpha\beta}\zeta^\beta \ ; \quad \zeta^\alpha = \zeta_\beta\epsilon^{\beta\alpha} \ ; \quad
    \bar\zeta_{\dot\alpha} = \epsilon_{\dot\alpha\dot\beta}\bar\zeta^{\dot\beta} \ ; \quad \bar\zeta^{\dot\alpha} = \bar\zeta_{\dot\beta}\epsilon^{\dot\beta\dot\alpha} \ ,
\end{align}
for any spinors $\zeta^\alpha,\bar\zeta^{\dot\alpha}$.

We define a basis vector $\hat t_{\alpha\dot\alpha} \equiv \del_{\alpha\dot\alpha}t$ along the $t$ axis (with respect to the flat metric $\eta_{ab}$, this is a constant, past-pointing unit vector). In addition, to define our lightcone ansatz, we fix an arbitrary left-handed spinor $q^\alpha$ (again, constant with respect to the flat $\eta_{ab}$). 

\subsection{The lightcone ansatz}

We are now ready to write our lightcone ansatz for a self-dual deformed metric over $dS_4$:
\begin{align}
    ds^2 &= \frac{1}{4}g_{\alpha\dot\alpha\beta\dot\beta}(x)dx^{\alpha\dot\alpha}dx^{\beta\dot\beta} \ ; \\
    g_{\alpha\dot\alpha\beta\dot\beta}(x) &= -\frac{2}{t^2}\epsilon_{\alpha\beta}\epsilon_{\dot\alpha\dot\beta} 
      - \frac{1}{t} q_\alpha q_\beta q_\gamma q_\delta \left(\del^\gamma{}_{\dot\alpha}\del^\delta{}_{\dot\beta} - \frac{2}{t}\hat t^\gamma{}_{(\dot\alpha}\del^\delta{}_{\dot\beta)}\right)\phi(x) \ . \label{eq:g_ansatz}
\end{align}
The first term in \eqref{eq:g_ansatz} is the pure $dS_4$ metric, while the second term is the self-dual deformation, generated by a scalar prepotential $\phi(x^a)$. We stress that eq. \eqref{eq:g_ansatz} is exact, rather than a linear approximation in $\phi(x)$. The prepotential $\phi(x)$ is subject to a field equation:
\begin{align}
   \Box\phi = -\frac{1}{8}q^\alpha q^\beta q^\gamma q^\delta 
     \big(t\del_{\alpha\dot\alpha}\del_{\beta\dot\beta}\phi - 4\hat t_{\alpha\dot\alpha}\del_{\beta\dot\beta}\phi\big)\big(\del_\gamma{}^{\dot\alpha}\del_\delta{}^{\dot\beta}\phi\big) \ , \label{eq:phi_equation}
\end{align}
where $\Box$ is the flat d'Alembertian:
\begin{align}
    \Box \equiv \eta^{ab}\frac{\del^2}{\del x^a\del x^b} = -\frac{1}{2}\del_{\alpha\dot\alpha}\del^{\alpha\dot\alpha} \ .
\end{align}
Eq. \eqref{eq:phi_equation} is again exact, consistent with the fact that self-dual GR has only cubic interactions \cite{Krasnov:2016emc}. The field equation \eqref{eq:phi_equation} can be extracted from Metsaev's (A)dS lightcone formalism \cite{Metsaev:2018xip}, but the actual metric ansatz \eqref{eq:g_ansatz} is, to our knowledge, new (see \cite{Akshay:2014pla} for some related work). It is instructive to compare eqs. \eqref{eq:g_ansatz}-\eqref{eq:phi_equation} to their counterparts in self-dual GR over Minkowski space \cite{Plebanski:1975wn,Siegel:1992wd,Monteiro:2022nqt}, i.e. to Plebanski's second ``heavenly equation'':
\begin{align}
     g_{\alpha\dot\alpha\beta\dot\beta}(x) = -2\epsilon_{\alpha\beta}\epsilon_{\dot\alpha\dot\beta} + q_\alpha q_\beta q^\gamma q^\delta\del_{\gamma\dot\alpha}\del_{\delta\dot\beta}\,\phi(x) \ ; \quad
     \Box\phi \sim q^\alpha q^\beta q^\gamma q^\delta \big(\del_{\alpha\dot\alpha}\del_{\beta\dot\beta}\phi\big)\big(\del_\gamma{}^{\dot\alpha}\del_\delta{}^{\dot\beta}\phi\big) \ . \label{eq:flat_self_dual}
\end{align}
We also list for comparison the corresponding ansatz and equation for self-dual Yang-Mills theory \cite{Bardeen:1995gk,Rosly:1996vr} (here, there is no distinction between Minkowski and de Sitter, since the classical theory is conformal):
\begin{align}
    A_{\alpha\dot\alpha}(x) = q_\alpha q^\beta\del_{\beta\dot\alpha}\,\phi(x) \ ; \quad \Box\phi \sim q^\alpha q^\beta \big(\del_{\alpha\dot\alpha}\phi\big) \big(\del_\beta{}^{\dot\alpha}\phi\big) \ , \label{eq:YM}
\end{align}
and for a conformally-massless scalar in $dS_4$ with cubic interaction:
\begin{align}
    \Phi(x) = \frac{1}{t}\phi(x) \ ; \quad \Box\phi \sim \frac{1}{t}\phi^2 \ . \label{eq:scalar}
\end{align}
We see that the de Sitter equations \eqref{eq:g_ansatz}-\eqref{eq:phi_equation} are just a (surprisingly simple) modification of the Minkowski ones \eqref{eq:flat_self_dual}, which in turn are the ``square'' of the Yang-Mills equations \eqref{eq:YM}.

Returning now to our field equation \eqref{eq:phi_equation}, we note that it can trivially be encoded as the variation of an action:
\begin{align}
     S = \frac{1}{32\pi G}\int d^4x\,\psi\left(\Box\phi + \frac{1}{8}q^\alpha q^\beta q^\gamma q^\delta
        \big(t\del_{\alpha\dot\alpha}\del_{\beta\dot\beta}\phi - 4\hat t_{\alpha\dot\alpha}\del_{\beta\dot\beta}\phi\big)\big(\del_\gamma{}^{\dot\alpha}\del_\delta{}^{\dot\beta}\phi\big) \right) \label{eq:S_lightcone}
\end{align}
with respect to the Lagrange multiplier $\psi(x)$. The overall factor in \eqref{eq:S_lightcone}, along with eqs. \eqref{eq:g_ansatz}-\eqref{eq:phi_equation} themselves, will be justified in section \ref{sec:Krasnov}. In fact, as we'll see, the action \eqref{eq:S_lightcone} is smarter than it appears. In particular, we'll see that $\psi(x)$ is actually a component of a linearized left-handed Weyl curvature perturbation:
\begin{align}
    \psi(x) = -\frac{1}{t^3}q^\alpha q^\beta q^\gamma q^\delta \Psi_{\alpha\beta\gamma\delta}(x) \ , \label{eq:psi}
\end{align}
where the left-handed Weyl tensor $\Psi_{\alpha\beta\gamma\delta}(x)$ is written in an orthonormal basis w.r.t. the curved metric \eqref{eq:g_ansatz}. Varying the action \eqref{eq:S_lightcone} with respect to $\phi(x)$, we obtain the field equation for $\psi(x)$ as:
\begin{align}
    \Box\psi = -\frac{1}{4}q^\alpha q^\beta q^\gamma q^\delta\left( 
      \big(t\del_{\alpha\dot\alpha}\del_{\beta\dot\beta}\phi - 2\hat t_{\alpha\dot\alpha}\del_{\beta\dot\beta}\phi\big)\big(\del_\gamma{}^{\dot\alpha}\del_\delta{}^{\dot\beta}\psi\big)
      + 4\big(\del_{\alpha\dot\alpha}\del_{\beta\dot\beta}\phi\big)\big(\hat t_\gamma{}^{\dot\alpha}\del_\delta{}^{\dot\beta}\psi\big) \right) \ . \label{eq:psi_equation}
\end{align}
This equation is linearized in $\psi(x)$, but exact in $\phi(x)$.

\subsection{Perturbation theory and ``almost-conservation'' of energy} \label{sec:ansatz:perturbation_theory}

The cubic action \eqref{eq:S_lightcone} and its field equations \eqref{eq:phi_equation},\eqref{eq:psi_equation} can be studied with the usual perturbative methods. Since we're working at the classical level, we will ignore the overall coefficient in front of the action.

At linear order, the field equations become simply $\Box\phi = \Box\psi = 0$, with the general solution given by a superposition of plane waves:
\begin{align}
    \phi^{(1)}(x) = \int_{k^2 = 0}\frac{d^3\mathbf{k}}{2\omega}\,a(k)\,e^{ik_a x^a} \ ; \quad \psi^{(1)}(x) = \int_{k^2 = 0}\frac{d^3\mathbf{k}}{2\omega}\,b(k)\,e^{ik_a x^a} \ . \label{eq:phi_psi_1} 
\end{align}
Here, $k_a = (-\omega,\mathbf{k})$ is a lightlike 4-momentum, $a(k_a)$ are mode coefficients, and the integration range is understood to include both positive and negative frequencies:
\begin{align}
    \int_{k^2 = 0} \equiv \int_{\omega=|\mathbf{k}|} + \int_{\omega=-|\mathbf{k}|} \ .
\end{align}
In spinor language, we can write the lightlike 4-momenta as:
\begin{align}
    k_a = -\frac{1}{2}\sigma_a^{\alpha\dot\alpha}k_{\alpha\dot\alpha} \ ; \quad k_{\alpha\dot\alpha} = \lambda_\alpha\tilde\lambda_{\dot\alpha} \ ; \quad 
    \tilde\lambda_{\dot\alpha} = \sign(\omega)\bar\lambda_{\dot\alpha} \ .
\end{align}
At higher orders in perturbation theory, we get off-shell propagators with non-lightlike 4-momenta $K_a$. Every such propagator converts a $\psi$-leg into a $\phi$-leg, and is associated with a factor of $\int d^4K \frac{1}{K^2}$, where $K^2\equiv\eta^{ab}K_a K_b$. Finally, we have a cubic vertex that couples two $\phi$-legs with 4-momenta $P_a,Q_a$ to a $\psi$-leg with 4-momentum $K_a$ (all with the outgoing sign convention):
\begin{align}
   \frac{1}{2i}q^\alpha q^\beta q_\gamma q_\delta P_{\alpha\dot\alpha}P_{\beta\dot\beta}Q^{\gamma\dot\alpha} \left(\hat t^{\delta\dot\beta} - \frac{1}{4}Q^{\delta\dot\beta}\frac{\del}{\del K_t} \right) 
     \delta^4(K_a + P_a + Q_a) + (P\leftrightarrow Q) \ . \label{eq:vertex}
\end{align} 
What is remarkable here is that, in addition to the conservation of spatial momentum $\mathbf{K+P+Q}$ (due to the spatial translation symmetry of the Poincare coordinates), we also have an ``almost-conservation'' of energy $K_t+P_t+Q_t$: the energy-conserving delta function is present in \eqref{eq:vertex}, and is merely acted on by a derivative. This happens because the background metric's $t$-dependence is expressed in \eqref{eq:phi_equation},\eqref{eq:S_lightcone},\eqref{eq:psi_equation} in a particularly ``mild'' way: as a factor of $t$ in front of one of the terms. To see that this behavior of self-dual GR is special, compare it e.g. to the cubic scalar case \eqref{eq:scalar}. There, we have a negative power of $t$ in the field equation, which translates into an \emph{integral} over energies, destroying energy conservation completely.

To close this section with an explicit example, we write out the second-order correction to the linearized solutions \eqref{eq:phi_psi_1}:
\begin{align}
    \begin{split}
      \phi^{(2)}(x) ={}& \frac{i}{2}\int d^4K\frac{1}{K^2}\,e^{iK_a x^a}\int_{k_1^2 = 0}\frac{d^3\mathbf{k}_1}{2\omega_1}\,a(k_1)\int_{k_2^2 = 0}\frac{d^3\mathbf{k}_2}{2\omega_2}\,a(k_2) \\
         &\times \langle q\lambda_1\rangle\langle q\lambda_2\rangle[\tilde\lambda_1\tilde\lambda_2] \left(\langle q\lambda_2\rangle\langle q\hat t\tilde\lambda_2]
          + \frac{1}{4} \langle q\lambda_1\rangle\langle q\lambda_2\rangle[\tilde\lambda_1\tilde\lambda_2]\frac{\del}{\del K_t} \right) \\ 
          &\times\delta^4(K - k_1 - k_2) \ ;
    \end{split} \\
    \begin{split}
      \psi^{(2)}(x) ={}& \frac{i}{2}\int d^4K\frac{1}{K^2}\,e^{iK_a x^a}\int_{k_1^2 = 0}\frac{d^3\mathbf{k}_1}{2\omega_1}\,a(k_1)\int_{k_2^2 = 0}\frac{d^3\mathbf{k}_2}{2\omega_2}\,b(k_2) \\
        &\times \langle q\lambda_1\rangle\langle q\lambda_2\rangle[\tilde\lambda_1\tilde\lambda_2]\left(\langle q\lambda_2\rangle \langle q\hat t\tilde\lambda_2] 
          + 2\langle q\lambda_1\rangle\langle q\hat t\tilde\lambda_1] 
         + \frac{1}{2} \langle q\lambda_1\rangle\langle q\lambda_2\rangle[\tilde\lambda_1\tilde\lambda_2]\frac{\del}{\del K_t} \right) \\ 
       &\times\delta^4(K - k_1 - k_2) \ ,
\end{split}
\end{align}
where we used shorthands for spinor inner products:
\begin{align}
    \langle q\lambda\rangle \equiv q_\alpha\lambda^\alpha \ ; \quad [\bar q\bar\lambda] \equiv \bar q_{\dot\alpha}\bar\lambda^{\dot\alpha} \ ; \quad 
    \langle q\hat t\bar\lambda] = q_\alpha t^\alpha{}_{\dot\alpha}\bar\lambda^{\dot\alpha} \ .
\end{align}

\section{Deriving the lightcone ansatz from Krasnov's formalism} \label{sec:Krasnov}

In this section, we derive our lightcone ansatz \eqref{eq:g_ansatz} and the associated action \eqref{eq:S_lightcone}, using Krasnov's chiral formulation of GR \cite{Krasnov:2011pp,Krasnov:2011up} (see important precursors in \cite{Capovilla:1989ac,Capovilla:1990qi}). In sections \ref{sec:Krasnov:review}-\ref{sec:Krasnov:self_dual}, we review the formulation itself. In section \ref{sec:Krasnov:ansatz}, we use it to study the lightcone ansatz. In the Appendix, we show that our ansatz is in fact the \emph{most general} solution of self-dual GR, in the sense that any perturbation away from it (that preserves the equations of self-dual GR) is pure gauge.

\subsection{The Krasnov formulation of GR} \label{sec:Krasnov:review}

The Krasnov formulation of GR makes crucial use of the spacetime dimension being $D=4$, and a nonzero cosmological constant $\Lambda\neq 0$ (however, see \cite{Krasnov:2021cva} for a formulation of the self-dual sector with $\Lambda=0$). As before, we normalize $\Lambda=3$, so as to have a unit de Sitter radius. As in the Cartan (vielbein) formulation, coordinate indices enter only via totally-antisymmetric differential forms. As in section \ref{sec:ansatz}, we denote coordinate indices by $(a,b,\dots)$ (these need not refer to Poincare coordinates, but later on they will). As in the Cartan formulation, we also use internal indices that refer to a flat tangent spacetime at every point. In the Cartan formulation, these are \emph{vector} indices, which can be decomposed into left-handed and right-handed spinor indices as usual. In contrast, in the Krasnov formulation, we use \emph{only left-handed spinor indices} for the internal flat spacetime. As before, we denote left-handed spinor indices by $(\alpha,\beta,\dots)$. In the present context, these refer \emph{not} to the Poincare coordinates of section \ref{sec:ansatz}, but to an orthonormal basis in the internal flat tangent spacetime. 

The fundamental variables in Krasnov's formulation are:
\begin{itemize}
    \item The left-handed half $\omega_a^{\alpha\beta} = \omega_a^{(\alpha\beta)}$ of the spin connection (a 1-form on the spacetime manifold).
    \item The left-handed half $\Psi^{\alpha\beta\gamma\delta} = \Psi^{(\alpha\beta\gamma\delta)}$ of the Weyl curvature tensor (a 0-form on the spacetime manifold). 
\end{itemize}
Crucially, neither the metric $g_{ab}$ nor the vielbein $e_a^{\alpha\dot\alpha}$ enter as fundamental variables.

From the connection $\omega_a^{\alpha\beta}$, we construct its curvature $F_{ab}^{\alpha\beta}$ in the usual way:
\begin{align}
    F_{ab}^{\alpha\beta} = 2\del_{[a}\omega_{b]}^{\alpha\beta} - \omega^\alpha{}_{\gamma[a}\,\omega^{\gamma\beta}{}_{b]} \ . \label{eq:F}
\end{align}
The left-handed Weyl tensor $\Psi^{\alpha\beta}_{\gamma\delta}$ can be viewed as a matrix over the 3d space of symmetric rank-2 left-handed spinors. Adding to this the identity matrix $\mathbb{1}^{\alpha\beta}_{\gamma\delta} \equiv \delta^\alpha_{(\gamma} \delta^\beta_{\delta)}$ and taking the matrix inverse, we obtain the geometric series:
\begin{align}
    \left(\left(\mathbb{1} - \tfrac{1}{2}\Psi\right)^{-1}\right)^{\alpha\beta}_{\gamma\delta} 
      = \delta^\alpha_{(\gamma} \delta^\beta_{\delta)} + \frac{1}{2}\Psi^{\alpha\beta}_{\gamma\delta} + O(\Psi^2) \ .
\end{align}
From these objects, the GR action can now be constructed as:
\begin{align}
    S = \frac{i}{128\pi G}\int d^4x\,\epsilon^{abcd} \left(\mathbb{1} - \tfrac{1}{2}\Psi\right)^{-1}_{\alpha\beta\gamma\delta}\,F_{ab}^{\alpha\beta}F_{cd}^{\gamma\delta} \ . \label{eq:S_full}
\end{align}
To relate this unusual formulation to the standard metric language, we define the chiral 2-form:
\begin{align}
    \Sigma_{ab}^{\alpha\beta} = \left(\left(\mathbb{1} - \tfrac{1}{2}\Psi\right)^{-1}\right)^{\alpha\beta}_{\gamma\delta} F_{ab}^{\gamma\delta} \ , \label{eq:Sigma}
\end{align}
which is related to the vielbein $e_a^{\alpha\dot\alpha}$ via:
\begin{align}
    \Sigma_{ab}^{\alpha\beta} = -e^\alpha{}_{\dot\alpha[a} e^{\beta\dot\alpha}_{b]} \label{eq:Sigma_e} \ .
\end{align}
There is no closed-form expression for $e_a^{\alpha\dot\alpha}$, since the formalism never fixes a frame for right-handed spinors. However, the metric $g_{ab} = -\frac{1}{2}e_{\alpha\dot\alpha a}e^{\alpha\dot\alpha}_b$ \emph{can} be derived in closed form, as \cite{Urbantke:1984eb}:
\begin{align}
    g_{ab} = \frac{\hat g_{ab}}{(-\det\hat g)^{1/6}} \ ; \quad 
    \hat g_{ab} = \frac{1}{24 i}\epsilon^{cdef} (\Sigma^\alpha{}_\beta)_{ac} (\Sigma^\beta{}_\gamma)_{bd} (\Sigma^\gamma{}_\alpha)_{ef} \ . \label{eq:Urbantke}
\end{align}

The remarkable statement \cite{Krasnov:2011pp,Krasnov:2011up,Krasnov:2016emc} is that, on solutions to the Euler-Lagrange equations of \eqref{eq:S_full}, this metric satisfies the vacuum Einstein equations with cosmological constant, and is consistent with the left-handed connection $\omega_a^{\alpha\beta}$ and Weyl curvature $\Psi^{\alpha\beta\gamma\delta}$.

\subsection{Self-dual sector and all-spinor indices} \label{sec:Krasnov:self_dual}

The full GR action \eqref{eq:S_full} can be treated as a perturbative expansion around the self-dual sector, by expanding in powers of the left-handed Weyl curvature $\Psi^{\alpha\beta\gamma\delta}$. The zeroth-order term is topological. Self-dual GR (along with linearized anti-self-dual perturbations) is contained in the first-order term:
\begin{align}
    S = \frac{i}{256\pi G}\int d^4x\,\epsilon^{abcd}\,\Psi_{\alpha\beta\gamma\delta}\,F_{ab}^{\alpha\beta}F_{cd}^{\gamma\delta} \ . \label{eq:S_self_dual}
\end{align}
At leading order in $\Psi^{\alpha\beta\gamma\delta}$, the chiral 2-form \eqref{eq:Sigma} is given by the curvature $F_{ab}^{\alpha\beta}$ itself, so the metric \eqref{eq:Urbantke} becomes:
\begin{align}
  g_{ab} = \frac{\hat g_{ab}}{(-\det\hat g)^{1/6}} \ ; \quad 
  \hat g_{ab} = \frac{1}{24 i}\epsilon^{cdef} (F^\alpha{}_\beta)_{ac} (F^\beta{}_\gamma)_{bd} (F^\gamma{}_\alpha)_{ef} \ . \label{eq:Urbantke_self_dual}
\end{align}
This metric now describes a \emph{self-dual} solution of GR with cosmological constant, with \emph{vanishing} left-handed Weyl curvature. The field $\Psi^{\alpha\beta\gamma\delta}$ then describes a linearized left-handed Weyl tensor, not derived from $g_{ab}$, but propagating on top of it. Using \eqref{eq:Sigma_e} and $\Sigma_{ab}^{\alpha\beta} = F_{ab}^{\alpha\beta}$, we can relate $\Psi^{\alpha\beta\gamma\delta}$ to its counterpart $c_{abcd}$ with coordinate indices as:
\begin{align}
    c_{abcd} = \frac{1}{16}F_{ab}^{\alpha\beta}F_{cd}^{\gamma\delta}\Psi_{\alpha\beta\gamma\delta} \ . \label{eq:c_Psi}
\end{align}
The field equation for the self-dual solution is obtained by varying \eqref{eq:S_self_dual} with respect to $\Psi^{\alpha\beta\gamma\delta}$:
\begin{align}
    F_{[ab}^{(\alpha\beta}F_{cd]}^{\gamma\delta)} = 0 \ . \label{eq:F_equation}
\end{align}
The field equation for the anti-self-dual perturbation is similarly obtained by varying with respect to $\omega_a^{\alpha\beta}$, but we won't need the full form of that equation here.

Now, let us re-introduce spinor notation $x^a\rightarrow x^{\alpha\dot\alpha}$ for the spacetime coordinates, as in \eqref{eq:spinors}. In doing so, we risk confusion between the curved coordinate indices and the flat internal spinor indices. To minimize such risk, we will always place the internal indices \emph{before} the coordinate ones, as in \eqref{eq:F},\eqref{eq:Urbantke},\eqref{eq:Urbantke_self_dual}. Raising and lowering of indices will always be performed with the flat spinor metrics $\epsilon_{\alpha\beta}$ and $\epsilon_{\dot\alpha\dot\beta}$, with no regard to the curved spacetime metric. Thus, $\omega^{\alpha\beta}_a$ and $F_{ab}^{\alpha\beta}$ become expressed as $\omega^{\alpha\beta}{}_{\gamma\dot\gamma} = \omega^{(\alpha\beta)}{}_{\gamma\dot\gamma}$, $F^{\alpha\beta}{}_{\gamma\delta} = F^{(\alpha\beta)}{}_{(\gamma\delta)}$ and $\tilde F^{\alpha\beta}{}_{\dot\gamma\dot\delta} = \tilde F^{(\alpha\beta)}{}_{(\dot\gamma\dot\delta)}$, where:
\begin{align}
   \omega^{\alpha\beta}_a = -\frac{1}{2}\sigma_a^{\gamma\dot\gamma}\omega^{\alpha\beta}{}_{\gamma\dot\gamma} \ ; \quad 
   F^{\alpha\beta}_{ab} = \frac{1}{4}\left(\sigma_a^{\gamma\dot\gamma}\sigma_b{}^{\delta}{}_{\dot\gamma}\,F^{\alpha\beta}{}_{\gamma\delta} 
     + \sigma_a^{\gamma\dot\gamma}\sigma_{b\gamma}{}^{\dot\delta} \tilde F^{\alpha\beta}{}_{\dot\gamma\dot\delta} \right) \ .
\end{align}
In these variables, eqs. \eqref{eq:F},\eqref{eq:S_self_dual},\eqref{eq:F_equation} become:
\begin{gather}
   F^{\alpha\beta}{}_{\gamma\delta} = \del_{(\gamma}{}^{\dot\gamma}\,\omega^{\alpha\beta}{}_{\delta)\dot\gamma} 
     - \frac{1}{2}\omega^\alpha{}_{\varepsilon(\gamma}{}^{\dot\gamma}\,\omega^{\varepsilon\beta}{}_{\delta)\dot\gamma} \ ; \quad
   \tilde F^{\alpha\beta}{}_{\dot\gamma\dot\delta} = -\del_{\gamma(\dot\gamma}\,\omega^{\alpha\beta\gamma}{}_{\dot\delta)} 
     + \frac{1}{2}\omega^\alpha{}_{\varepsilon\gamma(\dot\gamma}\,\omega^{\varepsilon\beta\gamma}{}_{\dot\delta)} \ ; \label{eq:F_spinor} \\
   S = \frac{1}{256\pi G}\int d^4x\,\Psi_{\alpha\beta\gamma\delta}
      (F^{\alpha\beta}{}_{\varepsilon\zeta}F^{\gamma\delta\varepsilon\zeta} - \tilde F^{\alpha\beta}{}_{\dot\varepsilon\dot\zeta}\tilde F^{\gamma\delta\dot\varepsilon\dot\zeta}) \ ; \label{eq:S_spinor} \\  
   F^{(\alpha\beta}{}_{\varepsilon\zeta}F^{\gamma\delta)\varepsilon\zeta} - \tilde F^{(\alpha\beta}{}_{\dot\varepsilon\dot\zeta}\tilde F^{\gamma\delta)\dot\varepsilon\dot\zeta} = 0 \ , \label{eq:F_equation_spinor}
\end{gather}
while the metric \eqref{eq:Urbantke_self_dual} becomes:
\begin{align}
  g_{\alpha\dot\alpha\beta\dot\beta} ={}& \frac{\hat g_{\alpha\dot\alpha\beta\dot\beta}}{(-\det\hat g_{ab})^{1/6}} \ ; \label{eq:Urbantke_spinor_g} \\
  \begin{split}
      \hat g_{\alpha\dot\alpha\beta\dot\beta} ={}& -\frac{1}{48}\left( \epsilon_{\dot\alpha\dot\beta}F^\gamma{}_{\delta\alpha\zeta} F^\delta{}_\varepsilon{}^\zeta{}_\xi F^\varepsilon{}_\gamma{}^\xi{}_\beta
        + 3F^\gamma{}_{\delta\alpha\zeta} F^\delta{}_\varepsilon{}^\zeta{}_\beta \tilde F^\varepsilon{}_{\gamma\dot\alpha\dot\beta} \right. \\
        &\left.\qquad {}- 3F^\gamma{}_{\delta\alpha\beta} \tilde F^\delta{}_{\varepsilon\dot\alpha\dot\gamma} \tilde F^\varepsilon{}_{\gamma}{}^{\dot\gamma}{}_{\dot\beta}
        - \epsilon_{\alpha\beta}\tilde F^\gamma{}_{\delta\dot\alpha\dot\gamma} \tilde F^\delta{}_\varepsilon{}^{\dot\gamma}{}_{\dot\delta} \tilde F^\varepsilon{}_\gamma{}^{\dot\delta}{}_{\dot\beta} \right) \ .
  \end{split} \label{eq:Urbantke_spinor_hat_g}
\end{align}
We didn't bother to write out $\det\hat g_{ab}$ in spinor form, since it will end up trivial in our application below. Finally, eq. \eqref{eq:c_Psi} can be translated into all-spinor indices as:
\begin{align}
  \begin{split}
    c_{abcd} ={}& \frac{1}{16}\left(\sigma_a^{\alpha\dot\alpha}\sigma_b{}^{\beta}{}_{\dot\alpha}\sigma_c^{\gamma\dot\gamma}\sigma_d{}^{\delta}{}_{\dot\gamma}\,c_{\alpha\beta\gamma\delta}
      + \left(\sigma_a^{\alpha\dot\alpha}\sigma_b{}^{\beta}{}_{\dot\alpha}\sigma_c^{\gamma\dot\gamma}\sigma_{d\gamma}{}^{\dot\delta} 
      + \sigma_a^{\gamma\dot\gamma}\sigma_{b\gamma}{}^{\dot\delta}\sigma_c^{\alpha\dot\alpha}\sigma_d{}^{\beta}{}_{\dot\alpha} \right) c_{\alpha\beta\dot\gamma\dot\delta} \right. \\
      &\qquad \left. {}+ \sigma_a^{\alpha\dot\alpha}\sigma_{b\alpha}{}^{\dot\beta}\sigma_c^{\gamma\dot\gamma}\sigma_{d\gamma}{}^{\dot\delta}\,c_{\dot\alpha\dot\beta\dot\gamma\dot\delta} \right) \ ; \\
    c_{\alpha\beta\gamma\delta} ={}& \frac{1}{16}F^{\varepsilon\zeta}{}_{\alpha\beta}F^{\xi\eta}{}_{\gamma\delta}\Psi_{\varepsilon\zeta\xi\eta} \ ; \\
    c_{\alpha\beta\dot\gamma\dot\delta} ={}& \frac{1}{16}F^{\varepsilon\zeta}{}_{\alpha\beta}\tilde F^{\gamma\delta}{}_{\dot\gamma\dot\delta}\Psi_{\varepsilon\zeta\gamma\delta} \ ; \\
    c_{\dot\alpha\dot\beta\dot\gamma\dot\delta} ={}& \frac{1}{16}\tilde F^{\alpha\beta}{}_{\dot\alpha\dot\beta}\tilde F^{\gamma\delta}{}_{\dot\gamma\dot\delta}\Psi_{\alpha\beta\gamma\delta} \ .
  \end{split} \label{eq:c_Psi_spinor}
\end{align}

\subsection{The lightcone ansatz} \label{sec:Krasnov:ansatz}

Having reviewed the Krasnov formulation, we are now ready to construct the lightcone ansatz of section \ref{sec:ansatz}. Following section \ref{sec:Krasnov:self_dual}, we will construct a connection $\omega^{\alpha\beta}{}_{\gamma\dot\gamma}$ for a self-dual solution, derive from it the solution's metric $g_{\alpha\dot\alpha\beta\dot\beta}$, and treat the anti-self-dual perturbation $\Psi^{\alpha\beta\gamma\delta}$ as a field propagating on this self-dual geometry.
    
Following \cite{Krasnov:2011up}, we separate $\omega^{\alpha\beta}{}_{\gamma\dot\gamma}$ into a background term $A^{\alpha\beta}{}_{\gamma\dot\gamma}$ that describes pure $dS_4$, plus a deformation $a^{\alpha\beta}{}_{\gamma\dot\gamma}$ (note that we're \emph{not} implying a linearized approximation):
\begin{align}
    \omega^{\alpha\beta}{}_{\gamma\dot\gamma} = A^{\alpha\beta}{}_{\gamma\dot\gamma} + a^{\alpha\beta}{}_{\gamma\dot\gamma} \ ; \quad \label{eq:omega}
    A^{\alpha\beta}{}_{\gamma\dot\gamma} = \frac{2}{t}\delta_\gamma^{(\alpha}\hat t^{\beta)}{}_{\dot\gamma} \ .
\end{align}
We then choose the following lightcone ansatz for the deformation $a^{\alpha\beta}{}_{\gamma\dot\gamma}$, inspired by the Yang-Mills ansatz \eqref{eq:YM}:
\begin{align}
    a^{\alpha\beta}{}_{\gamma\dot\gamma} = \frac{1}{t}q^\alpha q^\beta q_\gamma q^\delta \del_{\delta\dot\gamma}\phi \ . \label{eq:a_ansatz}
\end{align}
Here, $\phi(x)$ will end up as the right-handed prepotential from section \ref{sec:ansatz}. For any choice of $q^\alpha$, the ansatz \eqref{eq:a_ansatz} satisfies a pair of gauge conditions:
\begin{align}
    a_{\alpha\beta\gamma\dot\gamma} = a_{(\alpha\beta\gamma)\dot\gamma} \ ; \quad \del_{\gamma\dot\gamma}(t a^{\alpha\beta\gamma\dot\gamma}) = 0 \ .
\end{align}
These coincide with the gauge conditions introduced in \cite{Krasnov:2011up}, except that in \cite{Krasnov:2011up}, the factor of $t$ is absent, as a result of taking a high-energy (effectively, flat-spacetime) limit. 

The curvature \eqref{eq:F_spinor} of the connection \eqref{eq:omega} takes the schematic form $F = dA + AA + da + \{A,a\} + aa$. The term $aa$ quadratic in the deformation vanishes, since it involves a contraction of $q^\alpha$ with itself. We are left with:
\begin{align}
      F^{\alpha\beta}{}_{\gamma\delta} &= \frac{4}{t^2}\delta^\alpha_{(\gamma}\delta^\beta_{\delta)} - \frac{1}{t}q^\alpha q^\beta q_\gamma q_\delta\Box\phi
        + \frac{2}{t^2}\left( q_\gamma q_\delta q^{(\alpha}\hat t^{\beta)\dot\gamma} - q^\alpha q^\beta q_{(\gamma}\hat t_{\delta)}{}^{\dot\gamma} \right) q^\varepsilon\del_{\varepsilon\dot\gamma}\phi \ ; \label{eq:F_ansatz} \\
    \tilde F^{\alpha\beta}{}_{\dot\gamma\dot\delta} &= -\frac{1}{t}q^\alpha q^\beta q_\gamma q_\delta\left(\del^\gamma{}_{\dot\gamma}\del^\delta{}_{\dot\delta}\phi 
      - \frac{2}{t}t^\gamma{}_{(\dot\gamma}\del^\delta{}_{\dot\delta)}\phi \right) \ , \label{eq:tilde_F_ansatz}
\end{align}
where the $dS_4$ background is described by the first term in \eqref{eq:F_ansatz}. Let us now plug \eqref{eq:F_ansatz}-\eqref{eq:tilde_F_ansatz} into the LHS of the self-dual field equation \eqref{eq:F_equation_spinor}. The pure-$dS_4$ contribution vanishes (as it must, since pure $dS_4$ is a solution to the self-dual equation). We are thus left with terms linear and quadratic in $\phi$. Moreover, we find that the $\sim 1/t^4$ terms from the $FF$ and $\tilde F\tilde F$ pieces cancel. Overall, we get:
\begin{align}
  \begin{split}
    &F^{(\alpha\beta}{}_{\varepsilon\zeta}F^{\gamma\delta)\varepsilon\zeta} - \tilde F^{(\alpha\beta}{}_{\dot\varepsilon\dot\zeta}\tilde F^{\gamma\delta)\dot\varepsilon\dot\zeta} \\
    &\qquad = -\frac{1}{t^3}q^\alpha q^\beta q^\gamma q^\delta\left(8\Box\phi + q^\varepsilon q^\zeta q^\xi q^\eta
         \big(t\del_{\varepsilon\dot\alpha}\del_{\zeta\dot\beta}\phi - 4\hat t_{\varepsilon\dot\alpha}\del_{\zeta\dot\beta}\phi \big)\big(\del_\xi{}^{\dot\alpha}\del_\eta{}^{\dot\beta}\phi\big) \right) \ .
  \end{split} \label{eq:F_equation_ansatz}
\end{align}
We recognize this as the scalar equation \eqref{eq:phi_equation}, multiplied by an overall tensor factor $\sim q^\alpha q^\beta q^\gamma q^\delta/t^3$. The fact that we end up with one scalar equation for the scalar degree of freedom $\phi(x)$ confirms the consistency of our ansatz \eqref{eq:a_ansatz} for $a^{\alpha\beta}{}_{\gamma\dot\gamma}$. Plugging \eqref{eq:F_equation_ansatz} into \eqref{eq:S_spinor}, we recover the action \eqref{eq:S_lightcone} in terms of the right-handed degree of freedom $\phi(x)$ and the left-handed degree of freedom $\psi(x)$, defined via \eqref{eq:psi}.

Finally, let us derive the explicit metric of our self-dual solution. To do this, we plug \eqref{eq:F_ansatz}-\eqref{eq:tilde_F_ansatz} into \eqref{eq:Urbantke_spinor_hat_g}. It's easy to see that most of the terms vanish. Indeed, the free left-handed spinor indices in \eqref{eq:F_ansatz}-\eqref{eq:tilde_F_ansatz} are all packaged in factors of $q^\alpha$ and $\delta^\alpha_\beta$. Since the result $\hat g_{\alpha\dot\alpha\beta\dot\beta}$ has only 2 left-handed indices, this means that any product of terms with more than 2 factors of $q^\alpha$ must involve a contraction of $q^\alpha$ with itself, and therefore will vanish. Moreover, in the first (i.e. $FFF$) term in \eqref{eq:Urbantke_spinor_hat_g}, index symmetry requires the free indices to be arranged as $\epsilon_{\alpha\beta}$, which rules out \emph{any} factors of $q^\alpha$. We are thus left with only the $FFF$ and $FF\tilde F$ terms, and with only the pure-$dS_4$ term in each factor of $F$. The result for the ``densitized'' metric $\hat g_{\alpha\dot\alpha\beta\dot\beta}$ reads:
\begin{align}
    \hat g_{\alpha\dot\alpha\beta\dot\beta} = -\frac{2}{t^6}\epsilon_{\alpha\beta}\epsilon_{\dot\alpha\dot\beta} 
      - \frac{1}{t^5}q_\alpha q_\beta q_\gamma q_\delta \left(\del^\gamma{}_{\dot\alpha}\del^\delta{}_{\dot\beta} - \frac{2}{t}\hat t^\gamma{}_{(\dot\alpha}\del^\delta{}_{\dot\beta)}\right)\phi(x) \ . \label{eq:hat_g}
\end{align} 
Now, we note that the determinant of \eqref{eq:hat_g} only receives contributions from the first term. To see this, note that the second term has its free left-handed indices all packaged in factors of $q^\alpha$. As a result, it can contribute to the determinant only via contractions of $q^\alpha$ with itself, which vanish. Since the first term in \eqref{eq:hat_g} is just the pure $dS_4$ metric multiplied by $1/t^4$, we conclude that the determinant is:
\begin{align}
  \det\hat g_{ab} = -\frac{1}{t^{24}} \ . \label{eq:det_hat_g}
\end{align}
Plugging \eqref{eq:hat_g}-\eqref{eq:det_hat_g} into \eqref{eq:Urbantke_spinor_g}, we recover the ansatz \eqref{eq:g_ansatz} for the metric of the self-dual solution. We have thus established:
\begin{itemize}
 \item The relationship \eqref{eq:g_ansatz} between the right-handed degree of freedom $\phi(x)$ and a complete self-dual metric solution.
 \item The relationship \eqref{eq:psi} between the left-handed degree of freedom $\psi(x)$ and a linearized left-handed Weyl tensor.
 \item The lightcone-gauge action \eqref{eq:S_lightcone} that governs $\phi(x)$ and $\psi(x)$.
\end{itemize}
For completeness, we also present the vielbein associated with the metric \eqref{eq:g_ansatz} and spin connection \eqref{eq:omega}-\eqref{eq:a_ansatz}:
\begin{align}
    \begin{split}
        e_a^{\beta\dot\beta} &= -\frac{1}{2}\sigma_a^{\alpha\dot\alpha}e^{\beta\dot\beta}{}_{\alpha\dot\alpha} \ ; \\
        e^{\beta\dot\beta}{}_{\alpha\dot\alpha} &= -\frac{2}{t}\delta_\alpha^\beta\delta_{\dot\alpha}^{\dot\beta} 
        - \frac{1}{2}q_\alpha q^\beta q^\gamma q^\delta \left(\del_\delta{}^{\dot\beta}\del_{\gamma\dot\alpha} - \frac{2}{t}\hat t_\delta{}^{\dot\beta}\del_{\gamma\dot\alpha}\right)\phi(x) \ .
    \end{split} \label{eq:e}
\end{align}
It's easy to check that this squares into the metric \eqref{eq:g_ansatz} via $g_{ab} = -\frac{1}{2}e_{\gamma\dot\gamma a}e^{\gamma\dot\gamma}_b$, as well as into the 2-form \eqref{eq:F_ansatz} via $F_{ab}^{\alpha\beta} = \Sigma_{ab}^{\alpha\beta} = -e^\alpha{}_{\dot\alpha[a} e^{\beta\dot\alpha}_{b]}$. We found \eqref{eq:e} by first taking the ``naive'' square root of \eqref{eq:g_ansatz} (given by \eqref{eq:e} with the $\alpha\dot\alpha\leftrightarrow\beta\dot\beta$ index pairs symmetrized), and then looking for an internal left-handed rotation that would reproduce \eqref{eq:F_ansatz}. Note that, as everywhere in the Krasnov formalism, the freedom of \emph{right}-handed internal rotations remains. The explicit vielbein \eqref{eq:e} represents a particular (simple but not unique) choice of the internal right-handed frame.

\section{Extracting static-patch scattering data} \label{sec:together}

\subsection{Overview} \label{sec:together:overview}

Having justified the lightcone ansatz of section \ref{sec:ansatz}, we are now ready to discuss its relation to the static-patch scattering problem from section \ref{sec:static}. In section \ref{sec:ansatz}'s Poincare coordinates, the static patch's future horizon is simply the past lightcone $t=-|\mathbf{x}|$ of the origin -- a regular lightlike bulk hypersurface. On the other hand, the \emph{past} horizon is (the past half of) the Poincare coordinates' \emph{past lightlike infinity} ($t=-\infty$, $|\mathbf{x}|=\infty$, $-\infty<t+|\mathbf{x}|\leq 0$). As usual, a boundary at infinity (in this case, only in the coordinate sense) ultimately simplifies matters, but requires some extra care. 

We proceed as follows. In section \ref{sec:together:powers_of_u}, we discuss the behavior of $\phi(x)$ and $\psi(x)$ near the past horizon, translating from the (singular on the horizon) Poincare coordinates to (regular on the horizon) coordinates $(u,v,\mathbf{\hat r})$, defined via the pure-$dS_4$ coordinate transformation \eqref{eq:uv_from_Poincare}. Then, in section \ref{sec:together:initial_right}, we describe how our Poincare-patch solutions induce right-handed initial data on the past horizon, as defined in section \ref{sec:static:framework}. In particular, we show that the metric in $(u,v,\mathbf{\hat r})$ coordinates satisfies the appropriate constraints \eqref{eq:past_horizon_metric} on the past horizon, and also, with appropriate initial conditions for the prepotential $\phi(x)$, the constraints \eqref{eq:intersection_metric}-\eqref{eq:intersection_v_derivs} on the intersection 2-sphere. Next, in section \ref{sec:together:initial_left}, we describe how $\psi(x)$ induces left-handed initial data on the past horizon.

Finally, in section \ref{sec:together:final}, we describe how the Poincare-patch solutions induce final data on the future horizon. Here, the constraints \eqref{eq:future_horizon_metric_1} that ensure the lightlike and affine structure of the horizon's lightrays are generally \emph{not} satisfied in our $(u,v,\mathbf{\hat r})$ coordinates. However, there is \emph{one particular lightray} of the future horizon that remains undeformed from pure $dS_4$: the one aligned with the fixed spinor $q^\alpha$ that defines our lightcone gauge. This is good enough: we can probe the future horizon one lightray at a time, by varying $q^\alpha$ and thus changing the lightcone gauge. Note that the same situation occurred in the Yang-Mills case \cite{Albrychiewicz:2021ndv} (however, in that case we could handle the full theory, whereas here we're restricted to the self-dual sector).

As in \cite{David:2019mos,Albrychiewicz:2020ruh,Albrychiewicz:2021ndv}, we can mostly ignore the distinction between the geodesically complete horizons $(u=0,v\in\bbR),(u\in\bbR,v=0)$ and the ``halved'' horizons $(u\geq 0,v=0),(u=0,v\geq 0)$ that actually bound the static patch. In particular, we can arbitrarily extend the physical initial data at $v\geq 0$ into the $v<0$ range, and then just throw away the unphysical $u<0$ half of the final data. This is justified, at least in perturbation theory, by the hyperbolic (i.e. causal) structure of the linear term in the field equation \eqref{eq:phi_equation}. On the other hand, unlike in \cite{David:2019mos,Albrychiewicz:2020ruh,Albrychiewicz:2021ndv}, we will be forced to impose some constraints on the intersection 2-sphere $u=v=0$, so as to ensure eqs. \eqref{eq:intersection_metric}-\eqref{eq:intersection_v_derivs}.

\subsection{The behavior of $\phi(x)$ and $\psi(x)$ near $u=0$} \label{sec:together:powers_of_u}

\subsubsection{Coordinate transformations}

We begin by copying the coordinate relation \eqref{eq:uv_from_Poincare} as:
\begin{align}
    (u,v,\mathbf{\hat r}) = \left(-\frac{1}{t}, \frac{\mathbf{x}^2 - t^2}{t}, \frac{\mathbf{x}}{|\mathbf{x}|} \right) \ . \label{eq:uv_from_Poincare_again}
\end{align}
Taking derivatives, we get:
\begin{align}
  \frac{\del u}{\del x^a} = (u^2,\vec 0) \ ; \quad \frac{\del v}{\del x^a} = -2\left(1 - \frac{uv}{2}, \sqrt{1-uv}\,\mathbf{\hat r} \right) \ ; \quad
  \frac{\del\mathbf{\hat r}}{\del x^a} = \frac{u}{\sqrt{1-uv}}(0, \delta_{ij} - \hat r_i\hat r_j) \ .
\end{align}
Near the past horizon $u=0$, this becomes:
\begin{align}
  \frac{\del u}{\del x^a} = O(u^2) \ ; \quad \frac{\del v}{\del x^a} = -2\left(1, \mathbf{\hat r} \right) + O(u) \ ; \quad \frac{\del\mathbf{\hat r}}{\del x^a} = u(0, \delta_{ij} - \hat r_i\hat r_j) + O(u^2) \ . \label{eq:derivatives}
\end{align}
Now, the 2-sphere point $\mathbf{\hat r}$ and its tangent vectors $\mathbf{m},\mathbf{\bar m}$ can be parameterized in terms of spinors $\chi^\alpha,\bar\chi^{\dot\alpha}$ (normalized as $\hat t_{\alpha\dot\alpha}\chi^\alpha\bar\chi^{\dot\alpha} = -1$):
\begin{align}
    (\sigma_t - \mathbf{\hat r}\cdot\boldsymbol{\sigma})^{\alpha\dot\alpha} = 2\chi^\alpha\bar\chi^{\dot\alpha} \ ; \quad 
    \mathbf{m}\cdot\boldsymbol{\sigma}^{\alpha\dot\alpha} = \chi^\alpha\chi^\beta\hat t_\beta{}^{\dot\alpha} \ ; \quad 
    \mathbf{\bar m}\cdot\boldsymbol{\sigma}^{\alpha\dot\alpha} = \bar\chi^{\dot\alpha}\bar\chi^{\dot\beta}\hat t^\alpha{}_{\dot\beta} \ . \label{eq:r_m_spinors}
\end{align}
In terms of these spinors, the derivatives \eqref{eq:derivatives} near $u=0$ become:
\begin{align}
  \del_{\alpha\dot\alpha}u = O(u^2) \ ; \quad \del_{\alpha\dot\alpha}v = 4\chi_\alpha\bar\chi_{\dot\alpha} + O(u) \ ; \quad \del_{\alpha\dot\alpha}\mathbf{\hat r} = O(u) \ , \label{eq:gradients}
\end{align}
where, for the moment, we omitted the details of the $O(u)$ term in $\del_{\alpha\dot\alpha}\mathbf{\hat r}$. Rewriting $\mathbf{\hat r}$ in terms of $\chi_\alpha\bar\chi_{\dot\alpha}$ as in \eqref{eq:r_m_spinors}, this term reads:
\begin{align}
  \del_{\alpha\dot\alpha}(\chi_\beta\bar\chi_{\dot\beta}) = -u\chi^\gamma\bar\chi^{\dot\gamma}(\hat t_{\gamma\dot\alpha}\hat t_{\beta\dot\gamma}\chi_\alpha\bar\chi_{\dot\beta} + 
    \hat t_{\gamma\dot\beta}\hat t_{\alpha\dot\gamma}\chi_\beta\bar\chi_{\dot\alpha}) + O(u^2) \ . \label{eq:gradient_chi}
\end{align}

\subsubsection{Small-$u$ behavior of $\phi(x)$ and $\psi(x)$}

Now, consider the linearized solution $\phi^{(1)}(x)$ from \eqref{eq:phi_psi_1} for the right-handed degree of freedom $\phi(x)$. Recall that this $\phi^{(1)}(x)$ is just a superposition of lightlike plane waves in the Poincare coordinates. It is well-known (see e.g. \cite{Albrychiewicz:2020ruh}) that its rescaled version $u^{-1}\phi^{(1)}(x) = -t\phi^{(1)}(x)$ is a regular function of $(u,v,\mathbf{\hat r})$ near $u=0$; this is essentially the statement that a radiative field in flat spacetime can be expanded in positive integer powers of $1/r$. In particular, at leading order in $u$, we have:
\begin{align}
 \lim_{u\to 0} u^{-1}\phi^{(1)}(x) = \pi i \int_{-\infty}^\infty d\omega\,a(\omega,\omega\mathbf{\hat r})\,e^{-i\omega v/2} \ , \label{eq:phi_past_horizon_linearized}
\end{align}
where $a(k_a)$ are the mode coefficients from \eqref{eq:phi_psi_1}. 

We can now show, order by order in perturbation theory, that the non-linear corrections to $\phi(x)$ preserve the regularity of $u^{-1}\phi(x)$ at $u=0$. Using the conformal relation between the flat and de Sitter d'Alembertians:
\begin{align}
    \Box\phi  = \frac{1}{t^3}(\Box_{dS} - 2)(t\phi) = u^3(\Box_{dS} - 2)\frac{\phi}{u}
\end{align}
and the relation $\del_{\alpha\dot\alpha}u^{-1} = -\del_{\alpha\dot\alpha}t = -\hat t_{\alpha\dot\alpha}$, we rewrite the field equation \eqref{eq:phi_equation} as:
\begin{align}
    \begin{split}
        (\Box_{dS} - 2)\frac{\phi}{u} = \frac{1}{8}\,q^\alpha q^\beta q^\gamma q^\delta 
        &\left(\frac{1}{u^2}\left(\del_{\alpha\dot\alpha}\del_{\beta\dot\beta}\frac{\phi}{u}\right)\left(\del_\gamma{}^{\dot\alpha}\del_\delta{}^{\dot\beta}\frac{\phi}{u}\right)
        + \frac{8}{u}\hat t_{\alpha\dot\alpha}\left(\del_{\beta\dot\beta}\frac{\phi}{u}\right)\left(\del_\gamma{}^{\dot\alpha}\del_\delta{}^{\dot\beta}\frac{\phi}{u}\right) \right. \\
        &\left. \quad {}+ 8\hat t_{\alpha\dot\alpha}\hat t_{\beta\dot\beta}\,\frac{\phi}{u}\left(\del_\gamma{}^{\dot\alpha}\del_\delta{}^{\dot\beta}\frac{\phi}{u}\right)
        - 6\hat t_{\alpha\dot\alpha}\hat t_{\beta\dot\beta}\left(\del_\gamma{}^{\dot\beta}\frac{\phi}{u}\right)\left(\del_\delta{}^{\dot\alpha}\frac{\phi}{u}\right) \right) \ .
    \end{split} \label{eq:phi_equation_power_counting}
\end{align}
Now, assume by induction that $u^{-1}\phi$ is a regular function of $(u,v,\mathbf{\hat r})$ near $u=0$, up to some order in perturbation theory. Recall that near $u=0$, the Poincare-coordinate gradients $\del_{\alpha\dot\alpha}$ of the $(u,v,\mathbf{\hat r})$ coordinates scale as in \eqref{eq:gradients}. Therefore, the gradient of $u^{-1}\phi$ (at the given order in perturbation theory) behaves as:
\begin{align}
  \del_{\alpha\dot\alpha}\frac{\phi}{u} = 4\chi_\alpha\bar\chi_{\dot\alpha}\del_v\frac{\phi}{u} + uf_{\alpha\dot\alpha} \ , \label{eq:grad_phi}
\end{align}
where $f_{\alpha\dot\alpha}$ is a regular function of $(u,v,\mathbf{\hat r})$ near $u=0$. Taking another gradient, and using eq. \eqref{eq:gradient_chi} to take the gradient of $\chi_\alpha\bar\chi_{\dot\alpha}$ in \eqref{eq:grad_phi}, we get:
\begin{align}
 \begin{split}
   \del_{\alpha\dot\alpha}\del_{\beta\dot\beta}\frac{\phi}{u} ={}& 16\chi_\alpha\bar\chi_{\dot\alpha}\chi_\beta\bar\chi_{\dot\beta}\,\del_v^2\frac{\phi}{u} 
       - 4u\chi^\gamma\bar\chi^{\dot\gamma}\left(\hat t_{\gamma\dot\alpha}\hat t_{\beta\dot\gamma}\chi_\alpha\bar\chi_{\dot\beta} 
             + \hat t_{\gamma\dot\beta}\hat t_{\alpha\dot\gamma}\chi_\beta\bar\chi_{\dot\alpha} \right) \del_v\frac{\phi}{u} \\
       &+ 4u\chi_\alpha\bar\chi_{\dot\alpha}\,\del_v f_{\alpha\dot\alpha} + O(u^2) \ . \label{eq:grad_grad_phi}
 \end{split}
\end{align}
Now, using the fact that $\bar\chi_{\dot\alpha}$ contracted with itself gives zero, we can see that the contractions of derivatives that appear on the RHS of \eqref{eq:phi_equation_power_counting} satisfy:
\begin{align}
   \left(\del_{\beta\dot\beta}\frac{\phi}{u}\right)\left(\del_\gamma{}^{\dot\alpha}\del_\delta{}^{\dot\beta}\frac{\phi}{u}\right) = O(u) \ ; \quad \left(\del_{\alpha\dot\alpha}\del_{\beta\dot\beta}\frac{\phi}{u}\right)\left(\del_\gamma{}^{\dot\alpha}\del_\delta{}^{\dot\beta}\frac{\phi}{u}\right) = O(u^2) \ . \label{eq:grad_contractions}
\end{align}
This implies that the RHS of \eqref{eq:phi_equation_power_counting} is a regular function of $(u,v,\mathbf{\hat r})$ near $u=0$, despite the negative powers of $u$ that appear in it. We conclude that $u^{-1}\phi$ remains regular near $u=0$ at each successive order in perturbation theory, as desired. Moreover, if we use the \emph{retarded} propagator to invert the $\Box_{dS} - 2$ on the LHS of \eqref{eq:phi_equation_power_counting} (which is equivalent to using retarded propagators in the Poincare perturbation theory of section \ref{sec:ansatz:perturbation_theory}), then the linearized value \eqref{eq:phi_past_horizon_linearized} of $u^{-1}\phi$ at $u=0$ becomes the exact initial data for the full non-linear solution.

The same analysis can be applied to the left-handed degree of freedom $\psi(x)$, with its field equation \eqref{eq:psi_equation}. The upshot is that $u^{-1}\phi(x)$ and $u^{-1}\psi(x)$ are both regular functions of $(u,v,\mathbf{\hat r})$ near the past horizon $u=0$, with initial data given by:
\begin{align}
   \hat\phi(v,\mathbf{\hat r}) &\equiv \lim_{u\to 0} u^{-1}\phi(u,v,\mathbf{\hat r}) = \pi i \int_{-\infty}^\infty d\omega\,a(\omega,\omega\mathbf{\hat r})\,e^{-i\omega v/2} \ ; \label{eq:phi_past_horizon} \\
   \hat\psi(v,\mathbf{\hat r}) &\equiv \lim_{u\to 0} u^{-1}\psi(u,v,\mathbf{\hat r}) = \pi i \int_{-\infty}^\infty d\omega\,b(\omega,\omega\mathbf{\hat r})\,e^{-i\omega v/2} \ . \label{eq:psi_past_horizon}
\end{align}
A priori, the initial data \eqref{eq:phi_past_horizon}-\eqref{eq:psi_past_horizon} are arbitrary functions of the coordinates $(v,\mathbf{\hat r})$ on the past horizon. As we will see below, to satisfy the geometric constraints \eqref{eq:intersection_metric}-\eqref{eq:intersection_v_derivs} on the intersection 2-sphere $u=v=0$, we should set $\hat\phi(v,\mathbf{\hat r})$ and its first three $\del_v$ derivatives at $v=0$ to zero:
\begin{align}
   \left.\hat\phi(v,\mathbf{\hat r})\right|_{v=0} = \left.\del_v\hat\phi(v,\mathbf{\hat r})\right|_{v=0} = \left.\del_v^2\hat\phi(v,\mathbf{\hat r})\right|_{v=0} = \left.\del_v^3\hat\phi(v,\mathbf{\hat r})\right|_{v=0} = 0 \ .
   \label{eq:phi_constraints}
\end{align}

\subsubsection{Behavior of the d'Alembertian at $u=v=0$}

To conclude this subsection, we spell out a consequence of the first three constraints in \eqref{eq:phi_constraints}, which will prove useful below. Let us return to the field equation \eqref{eq:phi_equation_power_counting}. Using the $dS_4$ metric in the form \eqref{eq:uv}, we can express the d'Alembertian $\Box_{dS}$ at $u=v=0$ as:
\begin{align}
    \Box_{dS} = 4\del_u\del_v + \Box_{S_2} \ ,
\end{align}
where $\Box_{S_2}$ is the 2-sphere Laplacian. Plugging this into \eqref{eq:phi_equation_power_counting}, we get, at $u=v=0$:
\begin{align}
     &\del_u\del_v\frac{\phi}{u} = \frac{1}{4}(2 - \Box_{S_2})\frac{\phi}{u} + \frac{1}{32}\,q^\alpha q^\beta q^\gamma q^\delta \left(\frac{1}{u^2}\left(\del_{\alpha\dot\alpha}\del_{\beta\dot\beta}\frac{\phi}{u}\right)\left(\del_\gamma{}^{\dot\alpha}\del_\delta{}^{\dot\beta}\frac{\phi}{u}\right) \right. \label{eq:del_u_v_phi} \\
     &\quad \left. {}+ \frac{8}{u}\hat t_{\alpha\dot\alpha}\left(\del_{\beta\dot\beta}\frac{\phi}{u}\right)\left(\del_\gamma{}^{\dot\alpha}\del_\delta{}^{\dot\beta}\frac{\phi}{u}\right)
       + 8\hat t_{\alpha\dot\alpha}\hat t_{\beta\dot\beta}\,\frac{\phi}{u}\left(\del_\gamma{}^{\dot\alpha}\del_\delta{}^{\dot\beta}\frac{\phi}{u}\right)
       - 6\hat t_{\alpha\dot\alpha}\hat t_{\beta\dot\beta}\left(\del_\gamma{}^{\dot\beta}\frac{\phi}{u}\right)\left(\del_\delta{}^{\dot\alpha}\frac{\phi}{u}\right) \right) \ . \nonumber
\end{align}
Now, let us assume the first three constraints in \eqref{eq:phi_constraints}. These imply that $u^{-1}\phi$ and its first two $\del_v$ derivatives at $v=0$ scale as $O(u)$ at small $u$. Plugging this into \eqref{eq:grad_phi}-\eqref{eq:grad_grad_phi}, we get the following small-$u$ scaling at $v=0$ for the gradients \eqref{eq:grad_phi}-\eqref{eq:grad_grad_phi} and their contractions \eqref{eq:grad_contractions}:
\begin{gather}
    \frac{\phi}{u} = O(u) \ ; \quad \del_{\alpha\dot\alpha}\frac{\phi}{u} = O(u) \ ; \quad \del_{\alpha\dot\alpha}\del_{\beta\dot\beta}\frac{\phi}{u} = O(u) \ ; \label{eq:grad_at_v0} \\
\left(\del_{\alpha\dot\alpha}\del_{\beta\dot\beta}\frac{\phi}{u}\right)\left(\del_\gamma{}^{\dot\alpha}\del_\delta{}^{\dot\beta}\frac{\phi}{u}\right) = O(u^3) \ , \label{eq:grad_contractions_at_v0}
\end{gather}
where, to obtain the last relation, we again used the fact that the contraction of $\bar\chi_{\dot\alpha}$ with itself vanishes. Plugging \eqref{eq:grad_at_v0}-\eqref{eq:grad_contractions_at_v0} into the RHS of \eqref{eq:del_u_v_phi}, we obtain:
\begin{align}
   \lim_{u\to 0}\left.\del_u\del_v\frac{\phi}{u}\right|_{v=0} = 0 \ . \label{eq:del_u_v_phi_vanish}
\end{align}

\subsection{Right-handed initial data on the past horizon, and constraints on the intersection 2-sphere} \label{sec:together:initial_right}

Let us now invert the coordinate relation \eqref{eq:uv_from_Poincare_again}:
\begin{align}
   x^a = (t,\mathbf{x}) &= \frac{1}{u}\left(-1, \sqrt{1-uv}\,\mathbf{\hat r} \right) \ .    
\end{align}
The basis vectors $(\ell,m,\bar m,n)$ from \eqref{eq:basis} can now be written in Poincare coordinates as:
\begin{align}
    \begin{split}
        n^a \equiv \frac{\del x^a}{\del v} &= \left(0, -\frac{1}{2\sqrt{1-uv}}\,\mathbf{\hat r} \right) = -\frac{1}{2}(0, \mathbf{\hat r}) + O(u) \ ; \\
        \ell^a \equiv \frac{\del x^a}{\del u} &= \frac{1}{u^2}\left(1, -\frac{1-\frac{1}{2}uv}{\sqrt{1-uv}}\,\mathbf{\hat r} \right) = \frac{1}{u^2}(1,-\mathbf{\hat r}) + \frac{v^2}{4}n^a + O(u) \ ; \\
        m^a \equiv \mathbf{m}\cdot\frac{\del x^a}{\del\mathbf{\hat r}} &= \frac{1}{u}\left(0, \sqrt{1-uv}\,\mathbf{m} \right) = \frac{1}{u}(0,\mathbf{m}) + O(1) \ ; \\
        \bar m^a \equiv \mathbf{\bar m}\cdot\frac{\del x^a}{\del\mathbf{\hat r}} &= \frac{1}{u}\left(0, \sqrt{1-uv}\,\mathbf{\bar m} \right) = \frac{1}{u}(0,\mathbf{m}) + O(1) \ ,
    \end{split} \label{eq:basis_vectors}
\end{align}
where we present both the exact expressions and their expansion in powers of $u$ near the past horizon $u=0$. In spinor indices, eqs. \eqref{eq:basis_vectors} become:
\begin{gather}
    n^{\alpha\dot\alpha} = \chi^\alpha\bar\chi^{\dot\alpha} + \frac{1}{2}\hat t^{\alpha\dot\alpha} + O(u) \ ; \quad
    \ell^{\alpha\dot\alpha} = \frac{2}{u^2}\chi^\alpha\bar\chi^{\dot\alpha} + \frac{v^2}{4}n^{\alpha\dot\alpha} + O(u) \ ; \label{eq:basis_spinor_1} \\
    m^{\alpha\dot\alpha} = \frac{1}{u}\chi^\alpha\chi^\beta\hat t_\beta{}^{\dot\alpha} + O(1) \ ; \quad 
    \bar m^{\alpha\dot\alpha} = \frac{1}{u}\bar\chi^{\dot\alpha}\bar\chi^{\dot\beta}\hat t^\alpha{}_{\dot\beta} + O(1) \ . \label{eq:basis_spinor_2}
\end{gather}
Now, let us rewrite our self-dual deformed metric \eqref{eq:g_ansatz} in terms of the function $u^{-1}\phi$ which is regular at $u=0$:
\begin{align}
   g_{\alpha\dot\alpha\beta\dot\beta}(x) = u^2\left( -2\epsilon_{\alpha\beta}\epsilon_{\dot\alpha\dot\beta} + q_\alpha q_\beta q_\gamma q_\delta \left(
     \del^\gamma{}_{\dot\alpha}\del^\delta{}_{\dot\beta} + 4u\hat t^\gamma{}_{(\dot\alpha}\del^\delta{}_{\dot\beta)} + 4u^2\hat t^\gamma{}_{\dot\alpha}\hat t^\delta{}_{\dot\beta} \right)\frac{\phi(x)}{u} \right) \ . \label{eq:g_past_horizon_raw}
\end{align}
To obtain the metric components in the $(\ell,m,\bar m,n)$ basis, we should contract \eqref{eq:g_past_horizon_raw} with the basis vectors \eqref{eq:basis_spinor_1}-\eqref{eq:basis_spinor_2}, and express the result in terms of the initial data $u^{-1}\phi$ (c.f. \eqref{eq:phi_past_horizon}) and its derivatives $\del_u,\del_v,\del_{\mathbf{\hat r}}$ along these basis vectors. Using the small-$u$ relation \eqref{eq:gradients} between these derivatives and the Poincare-coordinate gradients $\del_{\alpha\dot\alpha}$, it's easy to evaluate 8 of the 10 metric components at $u=0$, along with the $\del_u$ derivative of the component $g_{nn}$:
\begin{gather}
   g_{nn} = g_{nm} = g_{n\bar m} = g_{\bar m\bar m} = g_{\ell\bar m} = 0 \ ; \quad g_{\ell n} = g_{m\bar m} = \frac{1}{2} \ ; \quad g_{mm} = 4\langle\chi q\rangle^4 \del_v^2\frac{\phi}{u} \ ; \label{eq:g_past_horizon_1} \\
   \del_u g_{nn} = 0 \ . \label{eq:g_past_horizon_2}
\end{gather}
The remaining components $g_{\ell\ell},g_{\ell m}$ at $u=0$ are more complicated. However, for all the components (including $g_{\ell\ell}$ and $g_{\ell m}$), we notice two features. First, the components are all finite or zero at $u=0$: negative powers of $u$ always come with a contraction of $\bar\chi_{\dot\alpha}$ with itself, which vanishes. Second, any $\del_u$ derivatives end up multiplied by positive powers of $u$, so they do not appear at $u=0$. Thus, all the components are expressed in terms of the initial data \eqref{eq:phi_past_horizon} on the past horizon (i.e. at $u=0$), and its derivatives $\del_v,\del_{\mathbf{r}}$ (at most two) along the horizon directions.

From \eqref{eq:g_past_horizon_1}-\eqref{eq:g_past_horizon_2}, we immediately see that the constraints \eqref{eq:past_horizon_metric} on the past horizon are satisfied. As we'll now show, the constraints \eqref{eq:intersection_metric}-\eqref{eq:intersection_v_derivs} on the intersection 2-sphere $u=v=0$ are satisfied as well, assuming the conditions \eqref{eq:phi_constraints} on our initial data. These do not follow directly from \eqref{eq:g_past_horizon_1}, because they involve also $g_{\ell\ell}$ and $g_{\ell m}$. Instead, we proceed as follows. First, the constraints are of course satisfied for the pure $dS_4$ metric $-2u^2\epsilon_{\alpha\beta}\epsilon_{\dot\alpha\dot\beta}$, as can be checked directly. Now, as discussed in the previous paragraph, the deviation of the metric components from their pure $dS_4$ values at $u=0$ are expressed in terms of the initial data $u^{-1}\phi$ and its (at most second) derivatives $\del_v,\del_{\mathbf{r}}$ along the past horizon. But the conditions \eqref{eq:phi_constraints} ensure that at $u=v=0$, the data $u^{-1}\phi$ and its $\del_v,\del_{\mathbf{r}}$ derivatives \emph{up to third order} all vanish. Therefore, the value and first $\del_v$ derivative of the metric components at $u=v=0$ are the same as in pure $dS_4$, thus ensuring the constraints \eqref{eq:intersection_metric},\eqref{eq:intersection_v_derivs}. 

It remains to check the constraints \eqref{eq:intersection_u_derivs} on the metric's $\del_u$ derivatives at $u=v=0$. For this purpose, the components computed in \eqref{eq:g_past_horizon_1} will be sufficient. For the pure $dS_4$ metric, the $\del_u$ derivatives in \eqref{eq:intersection_u_derivs} all vanish. The deformation from pure $dS_4$, governed by $u^{-1}\phi$, can then be analyzed as follows. Since $u^{-1}\phi$ and its first two $\del_v$ derivatives vanish at $u=v=0$, the small-$u$ behavior given in \eqref{eq:g_past_horizon_1} simply rolls over to the \emph{next order} in small $u$, i.e. to the $\del_u$ derivatives:
\begin{gather}
    \del_u g_{nm} = \del_u g_{n\bar m} = \del_u g_{\bar m\bar m} = \del_u g_{m\bar m} = 0 \ ; \\ 
    \del_u g_{mm} = 4\langle\chi q\rangle^4 \del_u\del_v^2\frac{\phi}{u} \ .
\end{gather}
The vanishing \eqref{eq:del_u_v_phi_vanish} of $\del_u\del_v(u^{-1}\phi)$ now implies that the remaining constraints \eqref{eq:intersection_u_derivs} at $u=v=0$ are indeed satisfied.

With the constraints on the past horizon and intersection 2-sphere verified, we proceed to identify the initial data on the past horizon, as defined in section \ref{sec:static:framework:horizons}. For the right-handed data, we use \eqref{eq:g_past_horizon_1} to write:
\begin{align}
    C_{nmnm}(v,\mathbf{\hat r}) = \del_v^2 g_{mm}(0,v,\mathbf{\hat r}) = 4\langle\chi q\rangle^4\del_v^4\hat\phi(v,\mathbf{\hat r}) \ . \label{eq:past_right_data}
\end{align}
Together with the constraints \eqref{eq:phi_constraints}, this uniquely determines the prepotential data $\hat\phi(v,\mathbf{\hat r})$ as a 4-fold definite integral of the Weyl curvature data $C_{nmnm}(v,\mathbf{\hat r})$:
\begin{align}
    \hat\phi(v,\mathbf{\hat r}) = \frac{1}{4\langle\chi q\rangle^4}\int_0^v dv'\int_0^{v'} dv''\int_0^{v''} dv'''\int_0^{v'''} dv''''\,C_{nmnm}(v'''',\mathbf{\hat r}) \ . \label{eq:initial_phi}
\end{align}

\subsection{Left-handed initial data on the past horizon} \label{sec:together:initial_left}

Let us now extract the \emph{left-}handed initial data $c_{n\bar mn\bar m}$ from the initial data \eqref{eq:psi_past_horizon} for $\psi(x)$. First, we'll need to convert the Weyl tensor $\Psi^{\alpha\beta\gamma\delta}$ from the internal spinor basis to the Poincare coordinate basis, via \eqref{eq:c_Psi_spinor}. Then we'll need to contract the left-handed Weyl tensor with two copies of the bivector $n\wedge\bar m$. In Poincare coordinates, at leading order in $u\to 0$, this bivector reads:
\begin{align}
    n^{\alpha\dot\alpha}\bar m^{\beta\dot\beta} - n^{\beta\dot\beta}\bar m^{\alpha\dot\alpha} = -\frac{1}{2u}\left(
       \hat t^\alpha{}_{\dot\gamma}\bar\chi^{\dot\gamma}\,\hat t^\beta{}_{\dot\delta}\bar\chi^{\dot\delta}\epsilon^{\dot\alpha\dot\beta} + \bar\chi^{\dot\alpha}\bar\chi^{\dot\beta}\epsilon^{\alpha\beta} \right) \ .
    \label{eq:nm}
\end{align}
Now, as a regular field in $dS_4$, the left-handed Weyl tensor should have finite components in the $(\ell,m,\bar m,n)$ basis, which is regular at the past horizon $u=0$. Therefore, it will have vanishing contraction with any bivector whose components in the $(\ell,m,\bar m,n)$ vanish at $u=0$. In particular, its contraction with $n\wedge\bar m$ shouldn't change upon shifting $n\wedge\bar m$ by any finite multiples of the bivectors $u(\ell\wedge n + m\wedge\bar m)$, $u^2(\ell\wedge m)$ or $u^2(\ell\wedge\bar m)$, which read: 
\begin{align}
  u(\ell^{\alpha\dot\alpha}n^{\beta\dot\beta} - \ell^{\beta\dot\beta}n^{\alpha\dot\alpha} + m^{\alpha\dot\alpha}\bar m^{\beta\dot\beta} - m^{\beta\dot\beta}\bar m^{\alpha\dot\alpha}) 
    &= \frac{2}{u}\chi^{(\alpha}\hat t^{\beta)}{}_{\dot\gamma}\bar\chi^{\dot\gamma}\epsilon^{\dot\alpha\dot\beta} \ ; \\
  u^2(\ell^{\alpha\dot\alpha}m^{\beta\dot\beta} - \ell^{\beta\dot\beta}m^{\alpha\dot\alpha}) &= -\frac{2}{u}\chi^\alpha\chi^\beta\epsilon^{\dot\alpha\dot\beta} \ ; \\
  u^2(\ell^{\alpha\dot\alpha}\bar m^{\beta\dot\beta} - \ell^{\beta\dot\beta}\bar m^{\alpha\dot\alpha}) &= -\frac{2}{u}\bar\chi^{\dot\alpha}\bar\chi^{\dot\beta}\epsilon^{\alpha\beta} \ .
\end{align}
By adding appropriate multiples of these bivectors, we can cancel the right-handed term in \eqref{eq:nm}, and tune the left-handed term to anything that has the same inner product with $\chi_\alpha\chi_\beta$. To make contact with eq. \eqref{eq:psi}, it will be particularly convenient to replace \eqref{eq:nm} in this way by a multiple of $q^\alpha q^\beta$:
\begin{align}
  n^{\alpha\dot\alpha}\bar m^{\beta\dot\beta} - n^{\beta\dot\beta}\bar m^{\alpha\dot\alpha} + \text{shifts} = -\frac{q^\alpha q^\beta\epsilon^{\dot\alpha\dot\beta}}{2u\langle\chi q\rangle^2} \ . \label{eq:shifted}
\end{align}
The actual shift terms in \eqref{eq:shifted} can be found by demanding that the shifted bivector have vanishing contraction with $q_\alpha$. Explicitly, they read:
\begin{align}
 \begin{split}
  \text{shifts} ={}& \frac{2\langle\bar\chi q\rangle u}{\langle\chi q\rangle}
                            (\ell^{\alpha\dot\alpha}n^{\beta\dot\beta} - \ell^{\beta\dot\beta}n^{\alpha\dot\alpha} + m^{\alpha\dot\alpha}\bar m^{\beta\dot\beta} - m^{\beta\dot\beta}\bar m^{\alpha\dot\alpha}) \\
    &+ \frac{\langle\bar\chi q\rangle^2 u^2}{\langle\chi q\rangle^2}(\ell^{\alpha\dot\alpha}m^{\beta\dot\beta} - \ell^{\beta\dot\beta}m^{\alpha\dot\alpha})
    - \frac{u^2}{4}(\ell^{\alpha\dot\alpha}\bar m^{\beta\dot\beta} - \ell^{\beta\dot\beta}\bar m^{\alpha\dot\alpha}) \ .
 \end{split}
\end{align}
Now, plugging the shifted bivector \eqref{eq:shifted} into \eqref{eq:c_Psi_spinor}, we see that we need only the ``fully left-handed part'' $F^{\alpha\beta}{}_{\gamma\delta}$ of the 2-form $F_{ab}^{\alpha\beta}$:
\begin{align}
  c_{n\bar mn\bar m} = \frac{q^\alpha q^\beta q^\gamma q^\delta}{16u^2\langle\chi q\rangle^4}\,c_{\alpha\beta\gamma\delta} \ ; \quad 
  c_{\alpha\beta\gamma\delta} = \frac{1}{16}F^{\varepsilon\zeta}{}_{\alpha\beta}F^{\xi\eta}{}_{\gamma\delta}\Psi_{\varepsilon\zeta\xi\eta} \ . \label{eq:c_component}
\end{align}
Analyzing our expression \eqref{eq:F_ansatz} for $F^{\alpha\beta}{}_{\gamma\delta}$ with the same power-counting methods as we used for the metric, we see that at leading order in $u\to 0$, it is given by the undeformed, pure-$dS_4$ value:
\begin{align}
  F^{\alpha\beta}{}_{\gamma\delta} = 4u^2\delta^\alpha_{(\gamma}\delta^\beta_{\delta)} \ .
\end{align}
We thus have simply $c_{\alpha\beta\gamma\delta} = u^4\Psi_{\alpha\beta\gamma\delta}$, and the left-handed Weyl component \eqref{eq:c_component} reads:
\begin{align}
  c_{n\bar mn\bar m} = \frac{u^2 q^\alpha q^\beta q^\gamma q^\delta}{16\langle\chi q\rangle^4}\,\Psi_{\alpha\beta\gamma\delta} \ .
\end{align}
Comparing with eq. \eqref{eq:psi}, we finally obtain the relation between the left-handed initial data on the past horizon and the initial data \eqref{eq:psi_past_horizon} for $\psi(x)$:
\begin{align}
  \hat\psi(v,\mathbf{\hat r}) = 16\langle\chi q\rangle^4 c_{n\bar mn\bar m}(v,\mathbf{\hat r}) \ .
\end{align}

\subsection{Future horizon data, one lightray at a time} \label{sec:together:final}

We now set out to extract the final data on the future horizon $v=0$. As mentioned in section \ref{sec:together:overview}, this becomes easy if we focus on a single one of the horizon's lightrays -- the one aligned with our gauge spinor $q^\alpha$. We can then vary $q^\alpha$, scanning through different lightcone gauges, and thus extract the data on any desired lightray. 

Let us then choose $q^\alpha$ normalized as $\hat t_{\alpha\dot\alpha}q^\alpha \bar q^{\dot\alpha} = -1$, and focus on the particular lightray (with respect to the pure $dS_4$ metric) parameterized by \eqref{eq:r_m_spinors} with $(\chi^\alpha,\bar\chi^{\dot\alpha})=(q^\alpha,\bar q^{\dot\alpha})$, i.e. defined by:
\begin{align}
  (v,\mathbf{\hat r}) = (0,\boldsymbol{\sigma}^{\alpha\dot\alpha}q_\alpha\bar q_{\dot\alpha}) \ . \label{eq:final_lightray}
\end{align}
It's easy to check that this \emph{remains} a lightray also in the deformed metric \eqref{eq:g_ansatz}, and that $u = -1/t$ remains an affine parameter along it. On this lightray, the basis vectors \eqref{eq:basis_vectors} read:
\begin{align}
    \ell^{\alpha\dot\alpha} = 2t^2 q^\alpha\bar q^{\dot\alpha} \ ; \quad m^{\alpha\dot\alpha} = -tq^\alpha q^\beta\hat t_\beta{}^{\dot\alpha} \ ; \quad 
    \bar m^{\alpha\dot\alpha} = -t\bar q^{\dot\alpha}\bar q^{\dot\beta}\hat t^\alpha{}_{\dot\beta} \ .
    \label{eq:basis_vectors_future}
\end{align}
It is now straightforward to extract the \emph{left-handed} final data $c_{\ell m\ell m}(u,\mathbf{\hat r})$ in terms of our left-handed degree of freedom $\psi(x)$. As in section \ref{sec:together:initial_left}, we plug into \eqref{eq:psi},\eqref{eq:c_Psi_spinor},\eqref{eq:F_ansatz} to translate between $c_{\ell m\ell m}$, $\Psi^{\alpha\beta\gamma\delta}$ and $\psi$, and we again find that only the pure-$dS_4$ part of the curvature \eqref{eq:F_ansatz} contributes. Altogether, the left-handed final data reads:
\begin{align}
    c_{\ell m\ell m}(u,\mathbf{\hat r}) = \frac{q^\alpha q^\beta q^\gamma q^\delta c_{\alpha\beta\gamma\delta}}{4u^6}
    = \frac{q^\alpha q^\beta q^\gamma q^\delta F^{\varepsilon\zeta}{}_{\alpha\beta}F^{\xi\eta}{}_{\gamma\delta}\Psi_{\varepsilon\zeta\xi\eta}}{64u^6} 
    = \frac{q^\alpha q^\beta q^\gamma q^\delta \Psi_{\alpha\beta\gamma\delta}}{4u^2} = \frac{\psi(x)}{4u^5} \ . \label{eq:c_final}
\end{align} 
With the \emph{right-handed} final data, we must be more careful. In particular, we can't directly use the formulas of section \ref{sec:static:framework}: those rely on coordinates adapted to the entire final horizon, whereas the lightcone ansatz is only adapted to the single lightray \eqref{eq:final_lightray}. Of course, we can always just directly compute the Weyl curvature of the metric \eqref{eq:g_ansatz}, and extract the appropriate component $C_{\ell\bar m\ell\bar m}$. An easier path is to note that the lightcone ansatz provides easy access to a \emph{different} null hypersurface contaning our lightray: the Poincare-coordinate ``null hyperplane'' $q_\alpha \bar q_{\dot\alpha}x^{\alpha\dot\alpha} = 0$. We can then extract $C_{\ell\bar m\ell\bar m}$ from the shear and expansion of this ``hyperplane''. This can be further simplified by recalling that the Weyl curvature is conformal. We can rescale the metric \eqref{eq:g_ansatz} to obtain a new metric $\check g_{ab} \equiv t^2g_{ab}$:
\begin{align}
    \check g_{\alpha\dot\alpha\beta\dot\beta} = -2\epsilon_{\alpha\beta}\epsilon_{\dot\alpha\dot\beta} 
    - q_\alpha q_\beta q_\gamma q_\delta \left(t\del^\gamma{}_{\dot\alpha}\del^\delta{}_{\dot\beta} - 2\hat t^\gamma{}_{(\dot\alpha}\del^\delta{}_{\dot\beta)}\right)\phi(x) \ . \label{eq:rescaled_g}
\end{align}
With this rescaled metric, the null hyperplane $q_\alpha \bar q_{\dot\alpha}x^{\alpha\dot\alpha} = 0$ has vanishing expansion and left-handed shear (just like the horizon in the original metric). We can define a vector basis on the hyperplane that plays the same role as the basis $(\ell,m,\bar m)$ on the horizon:
\begin{align}
    \check\ell^{\alpha\dot\alpha} = 2q^\alpha\bar q^{\dot\alpha} \ ; \quad \check m^{\alpha\dot\alpha} = q^\alpha q^\beta\hat t_\beta{}^{\dot\alpha} \ ; \quad 
    \check{\bar m}^{\alpha\dot\alpha} = \bar q^{\dot\alpha}\bar q^{\dot\beta}\hat t^\alpha{}_{\dot\beta} \ .
\end{align}
Specifically, these basis vectors Lie-commute, $\check\ell$ is an affine null generator with respect to the metric \eqref{eq:rescaled_g}, $\check m$ is a left-handed null vector, and the inner product of $\check m,\check{\bar m}$ under the metric \eqref{eq:rescaled_g} is $\check g_{m\bar m} = \frac{1}{2}$. We can now extract the Weyl component $\check C_{\ell\bar m\ell\bar m}$ for the metric \eqref{eq:rescaled_g} as:
\begin{align}
    \check g_{\bar m\bar m} &= g_{\bar m\bar m} = -\frac{1}{4}\left(t(q^\alpha\bar q^{\dot\alpha}\del_{\alpha\dot\alpha})^2 + 2q^\alpha\bar q^{\dot\alpha}\del_{\alpha\dot\alpha}\right)\phi(x) \ ; \label{eq:g_component_final} \\
    \check C_{\ell\bar m\ell\bar m} &= (\check\ell^a\del_a)^2\check g_{\bar m\bar m} = -\frac{t}{4}(q^\alpha\bar q^{\dot\alpha}\del_{\alpha\dot\alpha})^4\phi(x) \ .
\end{align}
Rescaling back into the metric \eqref{eq:g_ansatz} and basis \eqref{eq:basis_vectors_future}, we finally obtain the right-handed data on our chosen lightray of the future horizon as:
\begin{align}
    C_{\ell\bar m\ell\bar m} = t^4\check C_{\ell\bar m\ell\bar m} = \frac{(q^\alpha\bar q^{\dot\alpha}\del_{\alpha\dot\alpha})^4\phi(x)}{4u^5} \ .
\end{align}
We note the simplicity of this formula (as compared e.g. with the metric component \eqref{eq:g_component_final}), as well as its similarity to the initial right-handed data \eqref{eq:past_right_data} and the final left-handed data \eqref{eq:c_final}.

\section{Outlook} \label{sec:outlook}

In this paper, we developed an exact lightcone ansatz for self-dual GR in $dS_4$, along with linearized anti-self-dual perturbations on top of it. In addition, we carefully formulated the static-patch scattering problem for this sector of GR, and showed how its solution at tree-level is encoded within our lightcone ansatz. 

There are a number of avenues for future work. First, it would be interesting to compute loop corrections to our classical static-patch scattering. In fact, it's likely that this problem can be solved completely, since self-dual GR is 1-loop-exact \cite{Krasnov:2016emc}. Second, it would of course be valuable to go from self-dual (complex) GR to full (real) GR. However, as discussed in section \ref{sec:static}, we expect this to be difficult both technically and conceptually. 

A more straightforward future direction is to continue up the ladder of spins, and address the static-patch scattering problem for the self-dual sector \cite{Ponomarev:2016lrm,Skvortsov:2018jea,Skvortsov:2020wtf,Sharapov:2022awp,Didenko:2022qga} of higher-spin gravity \cite{Vasiliev:1990en,Vasiliev:1995dn,Vasiliev:1999ba}. The present work should be directly relevant for this purpose, given that self-dual higher-spin theory has recently been formulated \cite{Krasnov:2021nsq} in a manner closely analogous to Krasnov's formulation of GR. It would be especially interesting to try and make contact between static-patch scattering and the holographic description \cite{Anninos:2011ui} of higher-spin gravity in $dS_4$.

\section*{Acknowledgements}

I am grateful to Aritra Banerjee, Kirill Krasnov and Julian Lang for discussions. This work was supported by the Quantum Gravity Unit of the Okinawa Institute of Science and Technology Graduate University (OIST).

\appendix
\section{Generality of the lightcone ansatz} \label{app:generality}

\subsection{The problem statement}

In this Appendix, we discuss the generality of our lightcone ansatz \eqref{eq:omega}-\eqref{eq:a_ansatz} for the connection $\omega_a^{\alpha\beta}$. Our claim is that this ansatz is general up to gauge and diffeomorphisms, in the sense that \emph{perturbations} around \eqref{eq:omega}-\eqref{eq:a_ansatz} can be described (at least locally) as gauge transformations. Here, the deviation \eqref{eq:a_ansatz} from pure $dS_4$ can be large, i.e. non-perturbative. Our treatment will borrow some ideas from the analysis \cite{Krasnov:2011up} of perturbations over pure $dS_4$.

Thus, let us consider the self-dual GR solution \eqref{eq:omega}-\eqref{eq:a_ansatz} as a background, with vielbein $e_a^{\alpha\dot\alpha}$ (given explicitly by \eqref{eq:e}) and covariant derivative $\nabla_{\alpha\dot\alpha} \equiv e^a_{\alpha\dot\alpha}\nabla_a$. Here and in the rest of this Appendix, we use strictly \emph{internal} spinor indices. This means that the translation between spinor and coordinate indices is always done using the curved vielbein $e_a^{\alpha\dot\alpha}$, rather than using the ``flat'' Pauli matrices $\sigma_a^{\alpha\dot\alpha}$ as we've done in the main text.

Now, consider a general linear perturbation of $\omega_a^{\alpha\beta}$ over our background:
\begin{align}
    \delta\omega_a^{\alpha\beta} = -\frac{1}{2}e_a^{\gamma\dot\gamma}\delta\Phi^{\alpha\beta}{}_{\gamma\dot\gamma} \ , \label{eq:delta_omega}
\end{align}
where $\delta\Phi_{\alpha\beta\gamma\dot\gamma}$ has the index symmetry $\delta\Phi_{\alpha\beta\gamma\dot\gamma} = \delta\Phi_{(\alpha\beta)\gamma\dot\gamma}$.

The perturbation to the curvature $F_{ab}^{\alpha\beta}$ is $\delta F_{ab}^{\alpha\beta} = 2\nabla_{[a}\delta\omega^{\alpha\beta}_{b]}$. To preserve the field equation \eqref{eq:F_equation}, this must satisfy $F_{[ab}^{(\alpha\beta}\delta F_{cd]}^{\gamma\delta)} = 0$. Converting to spinor indices and recalling that $F_{ab}^{\alpha\beta}$ is equal to the chiral 2-form \eqref{eq:Sigma_e}, we obtain the field equation for the connection perturbation \eqref{eq:delta_omega} as:
\begin{align}
    \nabla_{(\alpha}{}^{\dot\delta} \delta\Phi_{\beta\gamma\delta)\dot\delta} = 0 \ . \label{eq:perturbation_field_eq_raw}
\end{align}
Some of the solutions to this equation are the linearized diffeomorphisms $\delta\omega_a^{\alpha\beta} = F_{ab}^{\alpha\beta}\xi^b$, or, in spinor indices:
\begin{align}
    \delta\Phi_{\alpha\beta\gamma\dot\gamma} = 2\epsilon_{\gamma(\alpha}\xi_{\beta)\dot\gamma} \ . \label{eq:diffeo}
\end{align}
Others are linearized internal left-handed rotations $\delta\omega_a^{\alpha\beta} = \nabla_a\theta^{\alpha\beta}$, or, in spinor indices:
\begin{align}
    \delta\Phi_{\alpha\beta\gamma\dot\gamma} = \nabla_{\gamma\dot\gamma}\theta_{\alpha\beta} \ , \label{eq:gauge_raw}
\end{align}
where $\theta^{\alpha\beta} = \theta^{(\alpha\beta)}$. Finally, we have solutions that preserve the ansatz \eqref{eq:a_ansatz}, with some perturbation $\delta\phi$ to the prepotential $\phi(x)$ that preserves its field equation \eqref{eq:phi_equation}. These take the form:
\begin{align}
    \delta\Phi_{\alpha\beta\gamma\dot\gamma} = \frac{1}{t}q_\alpha q_\beta q_\gamma q^\delta\nabla_{\delta\dot\gamma}\delta\phi \ . \label{eq:preserving_ansatz}
\end{align}
Note that in this case, the translation between ``coordinate'' spinor indices in \eqref{eq:a_ansatz} and ``internal'' ones in \eqref{eq:preserving_ansatz} ends up being trivial. This is because the deviation of the vielbein \eqref{eq:e} from pure $dS_4$ is always along $q^\alpha$, which vanishes when contracted with the $q^\alpha$ factors in \eqref{eq:a_ansatz}. Our task now is to show that together, eqs. \eqref{eq:diffeo}-\eqref{eq:preserving_ansatz} span (at least locally) the most general solution $\delta\Phi_{\alpha\beta\gamma\dot\gamma}(x)$ to the field equation \eqref{eq:perturbation_field_eq_raw}. 

\subsection{Index symmetries}

The first step is to recognize a remarkable simplifying feature \cite{Krasnov:2011up} of the Krasnov formalism. Algebraically, $\delta\Phi_{\alpha\beta\gamma\dot\gamma}$ consists of two pieces with different index symmetries:
\begin{align}
    \delta\Phi_{\alpha\beta\gamma\dot\gamma} = \delta\Phi_{(\alpha\beta\gamma)\dot\gamma} + \frac{2}{3}\epsilon_{\gamma(\alpha}\delta\Phi_{\beta)\delta}{}^\delta{}_{\dot\gamma} \ ,
\end{align}
where all possible values of the second piece are precisely spanned by the diffeomorphisms \eqref{eq:diffeo}. Thus, we can use diffeomorphisms to restrict to totally-symmetric $\delta\Phi_{\alpha\beta\gamma\dot\gamma} = \delta\Phi_{(\alpha\beta\gamma)\dot\gamma}$. The ansatz-preserving perturbation \eqref{eq:preserving_ansatz} is already of this form, while the internal gauge transformation \eqref{eq:gauge_raw} gets index-symmetrized (via an appropriate diffeomorphism) into:
\begin{align}
    \delta\Phi_{\alpha\beta\gamma}{}^{\dot\gamma} = \nabla_{(\gamma}{}^{\dot\gamma}\theta_{\alpha\beta)} \ . \label{eq:gauge}
\end{align}
Thus, our remaining task is to show that eqs. \eqref{eq:preserving_ansatz},\eqref{eq:gauge} span the most general totally-symmetric solution to \eqref{eq:perturbation_field_eq_raw}. 

Our next step is to \emph{un}-symmetrize the spinor indices in the field equation \eqref{eq:perturbation_field_eq_raw}. That is, we aim to bring the field equation to the stronger form:
\begin{align}
    \nabla_\alpha{}^{\dot\delta} \delta\Phi_{\beta\gamma\delta\dot\delta} = 0 \ . \label{eq:perturbation_field_eq}
\end{align}
This is equivalent to the original equation \eqref{eq:perturbation_field_eq_raw} together with a ``Lorentz gauge'' condition $\nabla_{\gamma\dot\gamma} \delta\Phi_{\alpha\beta}{}^{\gamma\dot\gamma} = 0$. We claim that this condition can be ensured by a gauge transformation of the form \eqref{eq:gauge}. To see this, we start by writing the divergence of \eqref{eq:gauge} as:
\begin{align}
  \nabla_{\gamma\dot\gamma} \delta\Phi_{\alpha\beta}{}^{\gamma\dot\gamma} 
     = \frac{1}{3}\nabla_{\gamma\dot\gamma}\left(\nabla^{\gamma\dot\gamma}\theta_{\alpha\beta} + 2\nabla_{(\alpha}{}^{\dot\gamma}\theta_{\beta)}{}^\gamma \right) \ .
\end{align}
In the second term, we decompose into symmetric and antisymmetric parts w.r.t. the spinor indices on the two derivatives, which gives:
\begin{align}
  \nabla_{\gamma\dot\gamma} \delta\Phi_{\alpha\beta}{}^{\gamma\dot\gamma} 
    = \frac{2}{3}\nabla_{\gamma\dot\gamma}\nabla^{\gamma\dot\gamma}\theta_{\alpha\beta} + \nabla_{(\alpha|\dot\gamma|}\nabla_\beta{}^{\dot\gamma}\theta_{\gamma)}{}^\gamma \ . \label{eq:gauge_divergence_raw}
\end{align}
Now, on a solution to self-dual GR, the only curvature component that contributes to the commutator $\nabla_{(\alpha}{}^{\dot\gamma}\nabla_{\beta)\dot\gamma}$ is the Ricci scalar, i.e. the cosmological constant:
\begin{align}
    \nabla_{(\alpha}{}^{\dot\gamma}\nabla_{\beta)\dot\gamma}\zeta^\gamma = \delta_{(\alpha}^\gamma\zeta_{\beta)} \ , \label{eq:LH_commutator}
\end{align}
for any left-handed spinor field $\zeta^\alpha$. Plugging \eqref{eq:LH_commutator} into \eqref{eq:gauge_divergence_raw}, we get:
\begin{align}
  \nabla_{\gamma\dot\gamma}\delta\Phi_{\alpha\beta}{}^{\gamma\dot\gamma} = \frac{2}{3}\left(\nabla_{\gamma\dot\gamma}\nabla^{\gamma\dot\gamma} - 2\right)\theta_{\alpha\beta} \ .
\end{align}
Thus, given any $\delta\Phi_{\alpha\beta\gamma\dot\gamma}$, we can always choose a gauge transformation \eqref{eq:gauge} by solving $\left(\nabla_{\gamma\dot\gamma}\nabla^{\gamma\dot\gamma} - 2\right)\theta_{\alpha\beta} = -\frac{3}{2}\nabla_{\gamma\dot\gamma}\delta\Phi_{\alpha\beta}{}^{\gamma\dot\gamma}$, so as to bring us into the ``Lorentz gauge'' $\nabla_{\gamma\dot\gamma}\delta\Phi_{\alpha\beta}{}^{\gamma\dot\gamma} = 0$. 

The upshot of this subsection is that the perturbation $\delta\Phi_{\alpha\beta\gamma\dot\gamma}$ can be restricted to be totally-symmetric $\delta\Phi_{\alpha\beta\gamma\dot\gamma} = \delta\Phi_{(\alpha\beta\gamma)\dot\gamma}$, and to satisfy the strengthened field equation \eqref{eq:perturbation_field_eq}. Our remaining task is to show that the solutions to \eqref{eq:perturbation_field_eq} are spanned by the ansatz-preserving \eqref{eq:preserving_ansatz} and the residual gauge transformations \eqref{eq:gauge}, where the gauge parameter is restricted to preserve ``Lorentz gauge'':
\begin{align}
    \left(\nabla_{\gamma\dot\gamma}\nabla^{\gamma\dot\gamma} - 2\right)\theta_{\alpha\beta} = 0 \ . \label{eq:Box_theta}
\end{align}

\subsection{Counting unconstrained derivatives}

Let us now show that eqs. \eqref{eq:preserving_ansatz},\eqref{eq:gauge} indeed span all solutions to the field equation \eqref{eq:perturbation_field_eq}. Our approach is to study the tower of derivatives around some fixed spacetime point $x$, and count the number of unconstrained components in these derivatives. In other words, at each order in the derivative expansion, we will count the components that aren't related, by field equations or constraints, to other components with equal or smaller number of derivatives. 

Before we begin, we note that as usual, the Dirac-like field equation \eqref{eq:perturbation_field_eq} implies a Klein-Gordon-like equation:
\begin{align}
   \nabla_{\alpha\dot\alpha}\nabla^{\alpha\dot\alpha}\delta\Phi_{\beta\gamma\delta\dot\delta} = \text{lower-derivative terms} \ . \label{eq:Klein_Gordon}
\end{align}
This follows from acting on \eqref{eq:perturbation_field_eq} with $\nabla^{\alpha\dot\alpha}$, and noting that $\nabla^{\alpha(\dot\alpha}\nabla_\alpha{}^{\dot\delta)}$ is a commutator of covariant derivatives, which reduces to a derivative-free curvature term.

Now, we're ready to count the unconstrained components of the $n$'th derivative of $\delta\Phi_{\alpha\beta\gamma\dot\gamma}$:
\begin{align}
    \nabla_{\alpha_1\dot\alpha_1}\dots\nabla_{\alpha_n\dot\alpha_n}\delta\Phi_{\beta\gamma\delta\dot\delta} \ . \label{eq:nth_grad_Phi}
\end{align}
Any commutator of two derivatives again results in a lower-derivative term. Together with eqs. \eqref{eq:perturbation_field_eq} and \eqref{eq:Klein_Gordon}, this implies that the only unconstrained components of \eqref{eq:nth_grad_Phi} are the ones totally symmetrized over the $n$ left-handed indices $(\alpha_1\dots\alpha_n)$, as well as over the $n+1$ right-handed indices $(\dot\alpha_1\dots\dot\alpha_n\dot\delta)$. Recalling that the 3 remaining indices $(\beta\gamma\delta)$ are also symmetrized, and that a rank-$k$ totally-symmetric spinor has $k+1$ independent components, we conclude that the number of unconstrained components in \eqref{eq:nth_grad_Phi} is $4(n+1)(n+2)$. 

Now, consider the ansatz-preserving perturbation \eqref{eq:preserving_ansatz}. Here, the $n$'th derivative of $\delta\Phi_{\alpha\beta\gamma\dot\gamma}$ is determined by the $n$'th derivative of $q^\alpha\nabla_{\alpha\dot\alpha}\delta\phi$:
\begin{align}
    \nabla_{\alpha_1\dot\alpha_1}\dots\nabla_{\alpha_n\dot\alpha_n}q^\beta\nabla_{\beta\dot\beta}\delta\phi \ , \label{eq:nth_grad_phi}
\end{align}
where the relevant constraint is a Klein-Gordon-like equation $\nabla_{\alpha\dot\alpha}\nabla^{\alpha\dot\alpha}\delta\phi = \dots$ that arises from \eqref{eq:phi_equation}. This leaves unconstrained the components of \eqref{eq:nth_grad_phi} that are totally symmetrized over the $n$ indices $(\alpha_1\dots\alpha_n)$, as well as over the $n+1$ indices $(\dot\alpha_1\dots\dot\alpha_n\dot\beta)$. Thus, the number of unconstrained components is (at least) $(n+1)(n+2)$.

It remains to show that the remaining $3(n+1)(n+2)$ unconstrained components of \eqref{eq:nth_grad_Phi} can be provided by the derivative expansion of the gauge transformation \eqref{eq:gauge}. To make sure that we're not double-counting, we contract \eqref{eq:gauge} with $q^\alpha$, which annihilates the ansatz-preserving solutions \eqref{eq:preserving_ansatz}. The $n$'th derivatives of the result then take the form:
\begin{align}
   \nabla_{\alpha_1\dot\alpha_1}\dots\nabla_{\alpha_n\dot\alpha_n}q^\beta\nabla_{(\beta|\dot\beta|}\theta_{\gamma\delta)} \ , \label{eq:nth_grad_theta}
\end{align}
where $\theta_{\alpha\beta}$ satisfies the Klein-Gordon-like constraint \eqref{eq:Box_theta}. This leaves unconstrained the components of \eqref{eq:nth_grad_theta} that are totally symmetrized over $(\alpha_1\dots\alpha_n)$ and $(\dot\alpha_1\dots\dot\alpha_n\dot\beta)$. Since the remaining free indices $(\gamma\delta)$ are also symmetrized, this makes (at least) $3(n+1)(n+2)$ unconstrained components, i.e. just enough to complete the most general solution to \eqref{eq:perturbation_field_eq}.


\begin{thebibliography}{99}

\bibitem{Krasnov:2011up}
K.~Krasnov,
``Gravity as a diffeomorphism invariant gauge theory,''
Phys. Rev. D \textbf{84}, 024034 (2011)
doi:10.1103/PhysRevD.84.024034
[arXiv:1101.4788 [hep-th]].

\bibitem{Krasnov:2011pp}
K.~Krasnov,
``Pure Connection Action Principle for General Relativity,''
Phys. Rev. Lett. \textbf{106}, 251103 (2011)
doi:10.1103/PhysRevLett.106.251103
[arXiv:1103.4498 [gr-qc]].

\bibitem{Metsaev:2018xip}
R.~R.~Metsaev,
``Light-cone gauge cubic interaction vertices for massless fields in AdS(4),''
Nucl. Phys. B \textbf{936}, 320-351 (2018)
doi:10.1016/j.nuclphysb.2018.09.021
[arXiv:1807.07542 [hep-th]].

\bibitem{Plebanski:1975wn}
J.~F.~Plebanski,
``Some solutions of complex Einstein equations,''
J. Math. Phys. \textbf{16}, 2395-2402 (1975)
doi:10.1063/1.522505

\bibitem{Przanowski:1983xpa}
M.~Przanowski,
``LOCALLY HERMITE EINSTEIN, SELFDUAL GRAVITATIONAL INSTANTONS,''
Acta Phys. Polon. B \textbf{14} (1983), 625-627

\bibitem{Lipstein:2023pih}
A.~Lipstein and S.~Nagy,
``Self-Dual Gravity and Color-Kinematics Duality in AdS4,''
Phys. Rev. Lett. \textbf{131}, no.8, 081501 (2023)
doi:10.1103/PhysRevLett.131.081501
[arXiv:2304.07141 [hep-th]].

\bibitem{Albrychiewicz:2021ndv}
E.~Albrychiewicz, Y.~Neiman and M.~Tsulaia,
``MHV amplitudes and BCFW recursion for Yang-Mills theory in the de Sitter static patch,''
JHEP \textbf{09}, 176 (2021)
doi:10.1007/JHEP09(2021)176
[arXiv:2105.07572 [hep-th]].

\bibitem{Boulware:1968zz}
D.~G.~Boulware and L.~S.~Brown,
``Tree Graphs and Classical Fields,''
Phys. Rev. \textbf{172}, 1628-1631 (1968)
doi:10.1103/PhysRev.172.1628

\bibitem{Anninos:2012qw}
D.~Anninos,
``De Sitter Musings,''
Int. J. Mod. Phys. A \textbf{27}, 1230013 (2012)
doi:10.1142/S0217751X1230013X
[arXiv:1205.3855 [hep-th]].

\bibitem{Halpern:2015zia}
I.~F.~Halpern and Y.~Neiman,
``Holography and quantum states in elliptic de Sitter space,''
JHEP \textbf{12}, 057 (2015)
doi:10.1007/JHEP12(2015)057
[arXiv:1509.05890 [hep-th]].

\bibitem{Bousso:2000nf}
R.~Bousso,
``Positive vacuum energy and the N bound,''
JHEP \textbf{11}, 038 (2000)
doi:10.1088/1126-6708/2000/11/038
[arXiv:hep-th/0010252 [hep-th]].

\bibitem{David:2019mos}
A.~David, N.~Fischer and Y.~Neiman,
``Spinor-helicity variables for cosmological horizons in de Sitter space,''
Phys. Rev. D \textbf{100}, no.4, 045005 (2019)
doi:10.1103/PhysRevD.100.045005
[arXiv:1906.01058 [hep-th]].

\bibitem{Albrychiewicz:2020ruh}
E.~Albrychiewicz and Y.~Neiman,
``Scattering in the static patch of de Sitter space,''
Phys. Rev. D \textbf{103}, no.6, 065014 (2021)
doi:10.1103/PhysRevD.103.065014
[arXiv:2012.13584 [hep-th]].

\bibitem{Hawking:1976jb}
S.~W.~Hawking,
``Gravitational Instantons,''
Phys. Lett. A \textbf{60}, 81 (1977)
doi:10.1016/0375-9601(77)90386-3

\bibitem{Bardeen:1995gk}
W.~A.~Bardeen,
``Selfdual Yang-Mills theory, integrability and multiparton amplitudes,''
Prog. Theor. Phys. Suppl. \textbf{123}, 1-8 (1996)
doi:10.1143/PTPS.123.1

\bibitem{Rosly:1996vr}
A.~A.~Rosly and K.~G.~Selivanov,
``On amplitudes in selfdual sector of Yang-Mills theory,''
Phys. Lett. B \textbf{399}, 135-140 (1997)
doi:10.1016/S0370-2693(97)00268-2
[arXiv:hep-th/9611101 [hep-th]].

\bibitem{Mason:2009afn}
L.~J.~Mason and D.~Skinner,
``Gravity, Twistors and the MHV Formalism,''
Commun. Math. Phys. \textbf{294}, 827-862 (2010)
doi:10.1007/s00220-009-0972-4
[arXiv:0808.3907 [hep-th]].

\bibitem{Bern:2017puu}
Z.~Bern, H.~H.~Chi, L.~Dixon and A.~Edison,
``Two-Loop Renormalization of Quantum Gravity Simplified,''
Phys. Rev. D \textbf{95}, no.4, 046013 (2017)
doi:10.1103/PhysRevD.95.046013
[arXiv:1701.02422 [hep-th]].

\bibitem{Sachs:1962zzb}
R.~K.~Sachs,
``On the Characteristic Initial Value Problem in Gravitational Theory,''
J. Math. Phys. \textbf{3}, no.5, 908-914 (1962)
doi:10.1063/1.1724305

\bibitem{Penrose:1980yx}
R.~Penrose,
``NULL HYPERSURFACE INITIAL DATA FOR CLASSICAL FIELDS OF ARBITRARY SPIN AND FOR GENERAL RELATIVITY,''
Gen. Rel. Grav. \textbf{12}, 225-264 (1980)
doi:10.1007/BF00756234

\bibitem{Reisenberger:2007ku}
M.~P.~Reisenberger,
``The Poisson bracket on free null initial data for gravity,''
Phys. Rev. Lett. \textbf{101}, 211101 (2008)
doi:10.1103/PhysRevLett.101.211101
[arXiv:0712.2541 [gr-qc]].

\bibitem{Krasnov:2016emc}
K.~Krasnov,
``Self-Dual Gravity,''
Class. Quant. Grav. \textbf{34}, no.9, 095001 (2017)
doi:10.1088/1361-6382/aa65e5
[arXiv:1610.01457 [hep-th]].

\bibitem{Akshay:2014pla}
Y.~S.~Akshay, S.~Ananth and M.~Mali,
``Light-cone gravity in AdS$_{4}$,''
Nucl. Phys. B \textbf{884}, 66-73 (2014)
doi:10.1016/j.nuclphysb.2014.04.015
[arXiv:1401.5933 [hep-th]].

\bibitem{Siegel:1992wd}
W.~Siegel,
``Selfdual N=8 supergravity as closed N=2 (N=4) strings,''
Phys. Rev. D \textbf{47}, 2504-2511 (1993)
doi:10.1103/PhysRevD.47.2504
[arXiv:hep-th/9207043 [hep-th]].

\bibitem{Monteiro:2022nqt}
R.~Monteiro, R.~Stark-Much\~ao and S.~Wikeley,
``Anomaly and double copy in quantum self-dual Yang-Mills and gravity,''
[arXiv:2211.12407 [hep-th]].

\bibitem{Capovilla:1989ac}
R.~Capovilla, T.~Jacobson and J.~Dell,
``General Relativity Without the Metric,''
Phys. Rev. Lett. \textbf{63}, 2325 (1989)
doi:10.1103/PhysRevLett.63.2325

\bibitem{Capovilla:1990qi}
R.~Capovilla, T.~Jacobson and J.~Dell,
``GRAVITATIONAL INSTANTONS AS SU(2) GAUGE FIELDS,''
Class. Quant. Grav. \textbf{7}, L1-L3 (1990)
doi:10.1088/0264-9381/7/1/001

\bibitem{Krasnov:2021cva}
K.~Krasnov and E.~Skvortsov,
``Flat self-dual gravity,''
JHEP \textbf{08}, 082 (2021)
doi:10.1007/JHEP08(2021)082
[arXiv:2106.01397 [hep-th]].

\bibitem{Urbantke:1984eb}
H.~Urbantke,
``ON INTEGRABILITY PROPERTIES OF SU(2) YANG-MILLS FIELDS. I. INFINITESIMAL PART,''
J. Math. Phys. \textbf{25}, no.7, 2321-2324 (1984)
doi:10.1063/1.526402

\bibitem{Ponomarev:2016lrm}
D.~Ponomarev and E.~D.~Skvortsov,
``Light-Front Higher-Spin Theories in Flat Space,''
J. Phys. A \textbf{50}, no.9, 095401 (2017)
doi:10.1088/1751-8121/aa56e7
[arXiv:1609.04655 [hep-th]].

\bibitem{Skvortsov:2018jea}
E.~D.~Skvortsov, T.~Tran and M.~Tsulaia,
``Quantum Chiral Higher Spin Gravity,''
Phys. Rev. Lett. \textbf{121}, no.3, 031601 (2018)
doi:10.1103/PhysRevLett.121.031601
[arXiv:1805.00048 [hep-th]].

\bibitem{Skvortsov:2020wtf}
E.~Skvortsov, T.~Tran and M.~Tsulaia,
``More on Quantum Chiral Higher Spin Gravity,''
Phys. Rev. D \textbf{101}, no.10, 106001 (2020)
doi:10.1103/PhysRevD.101.106001
[arXiv:2002.08487 [hep-th]].

\bibitem{Sharapov:2022awp}
A.~Sharapov and E.~Skvortsov,
``Chiral Higher Spin Gravity in (A)dS${}_4$ and secrets of Chern-Simons Matter Theories,''
[arXiv:2205.15293 [hep-th]].

\bibitem{Didenko:2022qga}
V.~E.~Didenko,
``On holomorphic sector of higher-spin theory,''
JHEP \textbf{10}, 191 (2022)
doi:10.1007/JHEP10(2022)191
[arXiv:2209.01966 [hep-th]].

\bibitem{Vasiliev:1990en}
M.~A.~Vasiliev,
``Consistent equation for interacting gauge fields of all spins in (3+1)-dimensions,''
Phys. Lett. B \textbf{243}, 378-382 (1990)
doi:10.1016/0370-2693(90)91400-6

\bibitem{Vasiliev:1995dn} 
M.~A.~Vasiliev,
``Higher spin gauge theories in four-dimensions, three-dimensions, and two-dimensions,''
Int.\ J.\ Mod.\ Phys.\ D {\bf 5}, 763 (1996)
[hep-th/9611024].

\bibitem{Vasiliev:1999ba} 
M.~A.~Vasiliev,
``Higher spin gauge theories: Star product and AdS space,''
In *Shifman, M.A. (ed.): The many faces of the superworld* 533-610
[hep-th/9910096].

\bibitem{Krasnov:2021nsq}
K.~Krasnov, E.~Skvortsov and T.~Tran,
``Actions for self-dual Higher Spin Gravities,''
JHEP \textbf{08}, 076 (2021)
doi:10.1007/JHEP08(2021)076
[arXiv:2105.12782 [hep-th]].

\bibitem{Anninos:2011ui} 
D.~Anninos, T.~Hartman and A.~Strominger,
``Higher Spin Realization of the dS/CFT Correspondence,''
Class.\ Quant.\ Grav.\  {\bf 34}, no. 1, 015009 (2017)
doi:10.1088/1361-6382/34/1/015009
[arXiv:1108.5735 [hep-th]].

\end{thebibliography}
\end{document}